\PassOptionsToPackage{dvipsnames}{xcolor}

\documentclass[conference]{IEEEtran}

\usepackage{todonotes}
\usepackage{textcomp}
\usepackage{gensymb}   
\usepackage{enumitem}  
\usepackage{hyperref}
\usepackage{multirow}
\usepackage{hhline}
\usepackage{float}
\usepackage{colortbl}
\usepackage{tikz} 
\usepackage{pgf}
\usepackage{pgfplots}  
\usepackage{url}
\usepackage[htt]{hyphenat}
\usepackage[T1]{fontenc}
\usepackage{lmodern}

\newcommand\rurl[1]{%
  \href{https://#1}{\nolinkurl{#1}}%
}

\usetikzlibrary{automata,positioning,backgrounds,patterns,arrows.meta}

\tikzset{
  every node/.style={font=\tiny},
  connector/.style={
      -stealth,
      font=\scriptsize
  },
  rectangle connector/.style={
      connector,
      to path={(\tikztostart) -- ++(#1,0pt) \tikztonodes |- (\tikztotarget) },
      pos=0.5
  }
}

\tikzstyle{wired} = [draw=blue, semithick]
\tikzstyle{wireless} = [draw=blue, dashed, semithick]
\tikzstyle{zigbee} = [draw=red, dashed, semithick]
\definecolor{a}{HTML}{d55e00}
\definecolor{b}{HTML}{cc79a7}
\definecolor{c}{HTML}{0072b2}
\definecolor{d}{HTML}{009e73}

\AtBeginDocument{%
  \providecommand\BibTeX{{%
    \normalfont B\kern-0.5em{\scshape i\kern-0.25em b}\kern-0.8em\TeX}}}

\begin{document}

\title{Supervising Smart Home Device Interactions: A Profile-Based Firewall Approach}


\author{\IEEEauthorblockN{François De Keersmaeker}
\IEEEauthorblockA{\textit{UCLouvain} \\
Louvain-la-Neuve, Belgium \\
francois.dekeersmaeker@uclouvain.be}
\and
\IEEEauthorblockN{Ramin Sadre}
\IEEEauthorblockA{\textit{UCLouvain} \\
Louvain-la-Neuve, Belgium \\
ramin.sadre@uclouvain.be}
\and
\IEEEauthorblockN{Cristel Pelsser}
\IEEEauthorblockA{\textit{UCLouvain} \\
Louvain-la-Neuve, Belgium \\
cristel.pelsser@uclouvain.be}
}

\maketitle

\begin{abstract}
Internet of Things devices can now be found everywhere, including in our households in the form of Smart Home deployments. Despite their ubiquity, their security is unsatisfactory, as demonstrated by recent attacks.

The IETF's MUD standard has as goal to simplify and automate the secure deployment of end devices in networks. A MUD file contains a device specific description of allowed network activities (e.g., allowed IP ports or host addresses) and can be used to configure for example a firewall. A major weakness of MUD is that it is not expressive enough to describe traffic patterns representing device interactions,
which often occur in modern Smart Home platforms.

In this article, we present a new language for describing such traffic patterns. The language allows writing device profiles that are more expressive than MUD files and take into account the interdependencies of traffic connections. We show how these profiles can be translated to efficient code for a lightweight firewall leveraging NFTables to block non-whitelisted traffic. We evaluate our approach on traffic generated by various Smart Home devices, and show that our system can accurately block unwanted traffic while inducing negligible latency.
\end{abstract}

\section{Introduction}

The success of the Internet of Thing (IoT) paradigm has led to a sharp increase in the use of networked embedded systems. On consumer level, a relevant subset of the IoT appears in Smart Homes. Those contain domestic objects, ranging from power plugs to washing machines, enhanced with sensing and communication capabilities. The market share of such devices has been estimated as \$99.89B in 2021, and is expected to grow exponentially in the next year, reaching \$380.52B in 2028 \cite{smart_home_market}.

Cyberattacks on Smart Home devices are a major concern for the privacy and security of their owners. Attackers could, for example, access live streams from IP cameras or remotely manipulate smart locks. Unfortunately, the current state of the security of IoT networks in general and Smart Home networks in particular is more than unsatisfactory. Due to resource and energy constraints, it is often difficult to run state-of-the-art security solutions, such as host-based intrusion detection systems, on IoT devices \cite{martins2022host}.
In addition, the devices are meant to be easy to use by people without network and cybersecurity knowledge, so it often depends entirely on the manufacturer whether and how often security updates are applied and best practices are followed. Finally, their large number makes it possible for attackers to misuse them to perform powerful attacks against other Internet hosts, as demonstrated by the Mirai botnet \cite{mirai}.

Improving the security of Smart Home devices has therefore become an urgency.
An assumption which is present in numerous research works
\cite{eskandari_passban_2020, pashamokhtari_progressive_2020, fu_hawatcher_2021}
is that an IoT device has a goal-oriented network behavior governed by a limited set of network access patterns that are simpler than what can be observed for general purpose devices such as computers and smartphones.
Based on this assumption, one can express the device's expected network behavior in a compact form that we will call a \emph{profile} in the following\footnote{The terms ``behavior'' and ``profile'' can be understood in different ways, but in the context of this paper, we will focus on the network communication issued from and toward the device.}.

Leveraging this concept, the Internet Engineering Task Force (IETF) defined the Manufacturer Usage Description (MUD) standard in RFC 8520 \cite{mud}. A MUD file is a machine-readable description of the network connections allowed for a device. It specifies one or more traffic access control lists based on typical traffic features, such as IP addresses, ports, protocols, connection direction, etc. More abstract traffic matching rules, e.g., matching on traffic originating from the same local network, are also supported.
The idea is to use the information contained in the MUD file of a device to configure local firewalls and switches such that the device can perform its normal functions (e.g., sending temperature measurements to a cloud) while blocking other activities.


A major weakness of MUD and similar approaches is the fact that they do not match the reality of today's Smart Home networks \cite{mazhar_role_2021}. In modern Smart Home deployments, devices can interact with each other directly or through one or more intermediate hosts. As a result, communication patterns emerge, consisting of temporally and logically interdependent connections. Such patterns cannot be described in MUD and, consequently, malicious activities breaking those patterns, e.g., an attacker directly accessing a device, cannot be blocked, which makes unauthorized device control and data exfiltration attacks possible \cite{ronen_extended_2016, uroz_characterization_2022}.
Two other, previously studied shortcomings are the lack of support for traffic statistics \cite{lear-bandwidth-2019} such as packet rate or count,
and the lack of supported protocols, mainly application-layer ones \cite{matheu_mud}.

In this article, we present a system to efficiently enforce legitimate device interactions in Smart Home networks. 
Our contributions are summarized as follows:

\textbf{We propose a new language to express profiles} that model legitimate network communication patterns of IoT devices. The language is more expressive than MUD profiles and takes into account the dependencies between traffic connections caused by device interactions.

\textbf{We describe a firewall} that leverages NFTables and NFQueue, lightweight firewall software available by default in Linux, to protect Smart Homes and a tool to automatically translate our device profiles into user-space code for the firewall.

\textbf{We evaluate our approach} on different types of Smart Home devices and show that our firewall correctly and efficiently forwards legitimate traffic while it blocks deviating traffic patterns with little overhead.

\textbf{We publish the source code} of our translation tool and firewall, the profiles that we wrote for the tested devices, and the traffic traces used in the evaluation,
at \texttt{https://github.com/franklin5168/smart-home-firewall} (pseudonym GitHub account).

On the other hand, \textbf{it is not the aim of this paper} to prove that the network communications of IoT devices have predictable properties that can be expressed in machine-readable descriptions, \textbf{nor} that such descriptions can be obtained with a reasonable effort.
These points have been discussed in the existing literature, to which we will refer in the relevant sections of this paper.

This article is organized as follows.
Section~\ref{sec:motivation} gives the motivation for our work.
In Section~\ref{sec:overview},
we present an overview of our system.
Section \ref{sec:system} then
delves more deeply into our system,
by describing our novel device network behavior profile specification,
our Smart Home stateful firewall implementation,
and our translator tool which generates code for the latter based on the former.
Section~\ref{sec:setup} presents the experimental setup for the evaluation of our approach, 
followed by Section~\ref{sec:evaluation} which presents our results.
Section~\ref{sec:discussion} discusses the strength and shortcomings of our system.
Section~\ref{sec:related-work} presents related works,
and the article concludes in Section~\ref{sec:conclusion}.

\section{Motivating example}
\label{sec:motivation}

This section will present the necessary background to understand our work,
as well as our motivations,
with the help of a concrete real-world example.
We will first briefly present Smart Home networks in general,
expose the concept of device interactions in such a network,
then describe potential attacks targeted toward such a network,
IETF's MUD standard and its shortcomings.
We will wrap up with a more formal definition of the threat model
considered in this article.

Our solution is intended to be deployed inside Smart Home networks,
to protect them from potentially malicious traffic.
The concept of Smart Homes involves heterogeneous everyday household objects,
from light bulbs to washing machines, enhanced with two new main capabilities:
They are equipped with sensors,
which make them able to measure data about their environment or their own operation.
And they possess a network interface,
such that they can share their measured data and can be remotely controlled.

Such ``smart'' objects are connected to the Local Area Network (LAN) of the home,
allowing them to communicate with each other,
with other, more traditional IT devices (e.g., laptops or smartphones),
or with hosts in the Internet via a router.
The ones capable to speak the TCP/IP protocols can be connected
through Wi-Fi or Ethernet.
Besides, others might use protocols specifically designed for Smart Home network communication,
the most widespread being Zigbee \cite{ergen2004zigbee}.
In that case, the devices need an intermediary,
which would translate between the two different protocols,
to be able to communicate with other devices;
such equipment is called a gateway.
The SmartThings system \cite{smartthings} is an example of such devices,
with end devices (e.g. power plug or door sensor) supporting only the Zigbee protocol,
and the hub serving as the gateway.

\begin{figure}
    \centering

    \tikzstyle{benign} = [-, draw=cyan]
    \tikzstyle{malicious} = [-Rays, draw=red]
    \begin{tikzpicture}[shorten >=1pt,semithick,>={stealth},node distance=2cm,on grid,auto]
        \node (door) [label={[yshift=0.15cm]below:{Door sensor}}] {\includegraphics[scale=0.05]{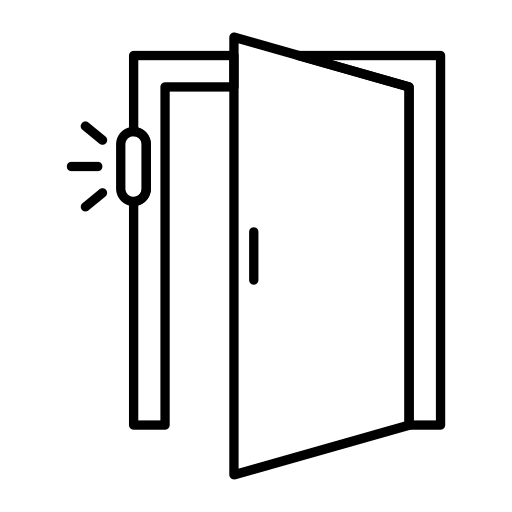}};
        \node (gateway) [label={[yshift=0.15cm]below:{Hub}}, right=2cm of door] {\includegraphics[scale=0.05]{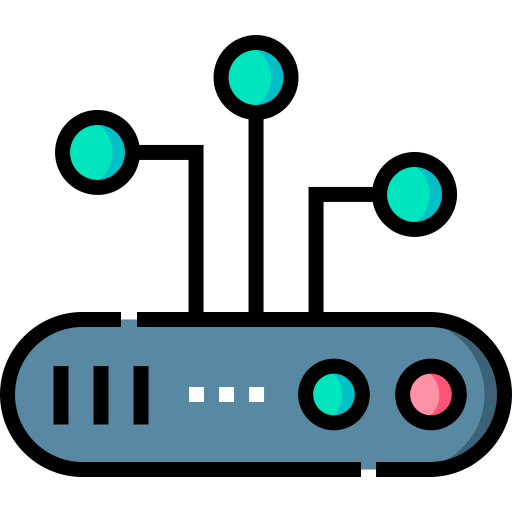}};
        \node (plug) [label={[yshift=0.15cm]below:{Smart plug}}, below=2cm of door] {\includegraphics[scale=0.05]{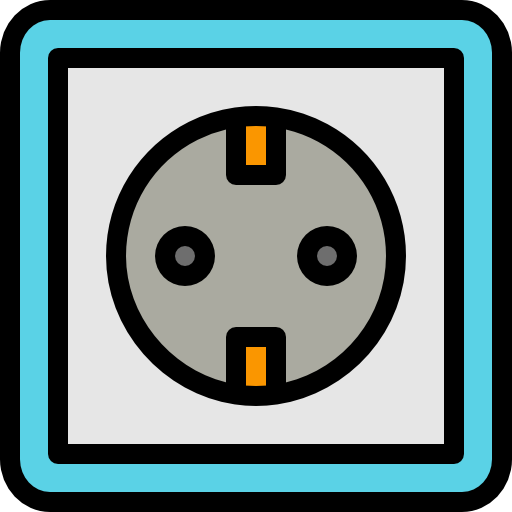}};
        \node (phone) [label={[yshift=0.15cm]below:{Smartphone}}, below=2cm of plug] {\includegraphics[scale=0.05]{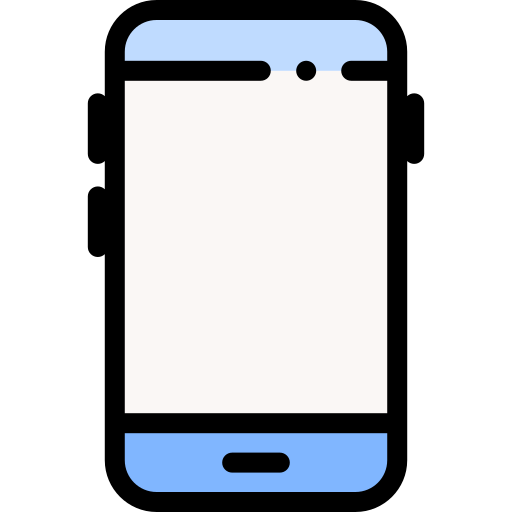}};
        \node (cpe) [label={[yshift=0.15cm]below:{CPE+AP}}, right=2cm of plug, xshift=2cm] {\includegraphics[scale=0.05]{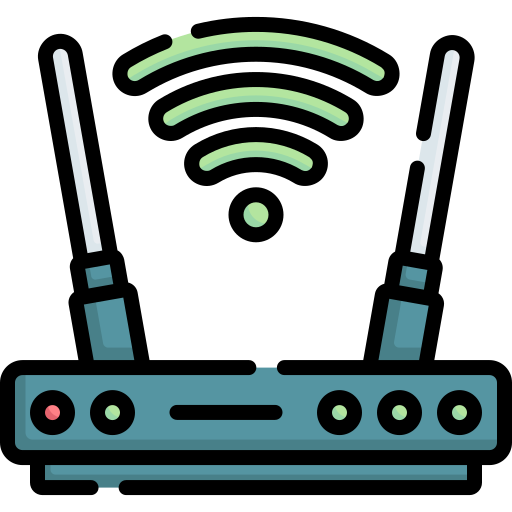}};
        \node (cloud) [label={[yshift=-0.45cm]above:{Internet}}, right=3cm of cpe] {\includegraphics[scale=0.1]{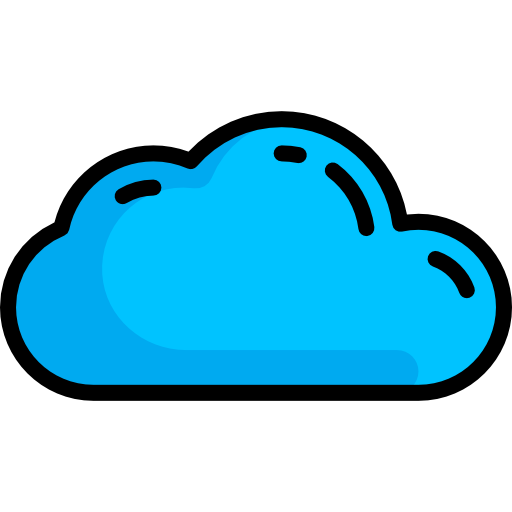}};
        \node (rogue-phone-lan) [label={[yshift=0.2cm,align=center]below:Rogue\\smartphone}, below=2cm of cpe] {\includegraphics[scale=0.05]{images/background/smartphone.png}};
        \node (rogue-phone-wan) [label={[yshift=0.2cm,align=center]below:Rogue\\smartphone}, right=3cm of rogue-phone-lan] {\includegraphics[scale=0.05]{images/background/smartphone.png}};
        \node (compromised-plug) [label={[align=center,yshift=0.15cm]below:{Compromised\\smart plug}}, above=2cm of cpe] {\includegraphics[scale=0.05]{images/background/power-socket.png}};

        \draw [benign, ->] (door) -- (gateway) node [midway, above, align=center] {\textbf{A1} Open};
        \draw [benign] (gateway) -- ([xshift=-0.01cm,yshift=0.1cm]cpe.center);
        \draw [benign, ->] ([xshift=0.025cm,yshift=0.1cm]cpe.center) -- ([yshift=0.09cm]cloud.west) node [midway, above, align=center] {\textbf{A2} Update};
        \draw [benign] ([xshift=0.025cm,yshift=-0.1cm]cpe.center) -- ([yshift=-0.09cm]cloud.west);
        \draw [benign, ->] ([xshift=0.025cm,yshift=-0.1cm]cpe.center) -- (plug) node [midway, above, align=center] {\textbf{A3} Command};
        \draw [benign] ([yshift=-0.2cm]cpe.west) -- (phone);
        \draw [benign, ->] ([yshift=-0.2cm]cpe.west) -- ([yshift=-0.1cm]plug.east) node [midway, below, align=left] {\textbf{B} ARP\\rate 1 pkt/s};
        \draw [malicious, -] (rogue-phone-wan) -- ([yshift=0.44cm]cloud.south);
        \draw [malicious] ([yshift=0.4cm]cloud.south) -- ([yshift=-0.45cm]cpe.east) node [midway, below, align=center] {\textbf{A} Command};
        \draw [malicious] (rogue-phone-lan) -- (cpe) node [midway, right, align=left, yshift=-0.1cm] {\textbf{A} Command\\\textbf{B} ARP rate > 1 pkt/s};
        \draw [malicious] (compromised-plug) -- (cpe) node [midway, right, align=left] {\textbf{C} DNS\\to C\&C};

        \begin{scope}[on background layer]

            \node (home-bottom-left) [below=0.9cm of phone, xshift=-0.8cm]  {};
            \node (home-top-right)   [above=1cm of compromised-plug, xshift=0.8cm] {};
            \fill[rounded corners, blue!10] (home-bottom-left) rectangle (home-top-right) {};
            \node (home-legend) [left=5.7cm of home-top-right, anchor=north west, xshift=0.1cm] {\footnotesize \textbf{Smart Home}};

        \end{scope}
    \end{tikzpicture}
    \caption{
        Motivating attack example. CPE: Customer Premises Equipment. AP: Access point.
    }
    \label{fig:example}
\end{figure}

A concrete example of such a deployment is depicted on Figure \ref{fig:example}.
It involves the following components:
\begin{itemize}
    \item a door sensor, using Zigbee to transmit its state to a SmartThings hub acting as a gateway;
    \item a smart plug, which communicates directly over TCP/IP via Wi-Fi;
    \item a smartphone, used to control the Smart Home devices from the local network, e.g. using their respective companion app;
    \item a router, which is the customer premises equipment (CPE) supplied by the ISP, and therefore acts as the network's access point and gateway toward the Internet.
\end{itemize}

\subsection{Device interactions}
\label{sec:device-interactions}

In the context of Smart Home networks,
device interactions are common,
and can even be considered as a core principle of Smart Home functionality.
Indeed, the main interest of a Smart Home environment is home automation,
which is achieved by having various devices seamlessly communicating with each other
to perform household tasks.
Based on our Smart Home network example described earlier (Figure \ref{fig:example}),
we consider a simple device interaction,
configured by the Smart Home user:
When the door sensor detects that the door opens (resp. closes),
the smart plug turns on (resp. off).
The network traffic related to each step of this interaction can be summarized as follows:
\begin{itemize}
    \item[\textbf{A1}] The door sensor indicates a change in the door's state, either open or close. It communicates using Zigbee.
    \item[\textbf{A2}] The hub translates the Zigbee message into TCP/IP, and forwards the change to the vendor's cloud.
    \item[\textbf{A3}] The cloud processes the message, and triggers the interaction by issuing a command toward the smart plug.
\end{itemize}

In this interaction, the action which actually toggled the plug is the third one,
i.e. the communication between the cloud and the plug itself.
However, this activity could not have occurred without
the previous communication between the hub and the cloud.

Obviously, the network activities depicted in the above examples carry a certain semantic and therefore happen in a specific order and between specific hosts. While a traditional firewall allows all these interactions, any time and in any possible order, \textbf{a smart firewall aware of the set of interactions, as proposed in this paper}, can block communications that break the expected patterns.
In this scenario, such an interaction-aware firewall would prevent, for example, a hacked smartphone to toggle the smart plug without being first triggered by the door sensor.

\subsection{Example of Attacks \& MUD Shortcomings}
\label{sec:attack-examples}

Taking the network setting presented on Figure~\ref{fig:example},
we will identify three attacks that cannot be protected against by using plain MUD,
whereas intended to be detected and blocked by our system.
Each attack will showcase a different added capability with respect to basic MUD,
namely the principle of traffic pattern interactions, traffic statistics, and matching for new, higher-layer protocols.

The first attack showcases the principle of interactions.
The intended behavior consists of the network patterns occurring
upon execution of the device interaction mentioned in section \ref{sec:device-interactions}:
the smart plug can be toggled by a packet coming from the vendor's cloud,
provided a door state change message has been transmitted by the hub beforehand.
As MUD does not support the interaction mechanism, or timing between packets,
a security solution leveraging MUD would accept all the packets necessary to allow this interaction,
regardless of their order.
However, our system will enforce the necessity of a preliminary door state change pattern,
before allowing the plug toggling pattern.
Any scenario where the plug would be toggled without a door state change pattern detected beforehand, would therefore be blocked.
This includes the following cases:
\begin{itemize}
    \item a remote attacker issuing an unwanted command toward the plug, from a distant network;
    \item a local attacker issuing an unwanted command toward the plug, from the local network;
    \item any unwanted traffic between the plug and its vendor's cloud, e.g. to exfiltrate privacy-sensitive data.
\end{itemize}

The second attack highlights the added traffic statistics feature.
We consider a simple intended behavior,
allowing pairs of ARP request/response messages between a local device
(e.g. a smartphone running the plug's companion app)
and the plug,
with a rate limited to one request per second,
illustrated by pattern \textbf{B} in Figure~\ref{fig:example}.
An attacker controlling a rogue device in the Smart Home LAN
will fail in trying to flood the network with ARP packets,
as all ARP the specified rate will be dropped.
This rate limit is unsupported by plain MUD,
but has already been studied by extensions to MUD \cite{lear-bandwidth-2019, singh_clearer_2019}.

Finally, we illustrate the newly added protocols matching capabilities with the third attack.
We take the DNS protocol as example.
It is intended for the smart plug to issue DNS requests and receive DNS responses
concerning its manufacturer's domain name(s).
We then assume that an attacker somehow managed to infect the smart plug,
and that they want to contact a malicious C\&C server located on the Internet.
The attacker will instruct the device to issue a DNS query towards the server's domain name.
Such traffic will be accepted if the security is configured using plain MUD,
as the DNS messages are only seen as UDP packets towards the DNS server's port 53,
which must be accepted to retain correct functionality.
Our system will nevertheless block such queries,
as it can read the specified domain name,
and block packets containing unintended names.

The two first attacks will be further instrumented and analyzed in Section \ref{sec:evaluation}.

\subsection{Threat model}
\label{sec:threat-model}

Based on the aforementioned attack examples,
we now define a more formal threat model
our system is intended to protect against.

Our focus is on attacks that manifest themselves as changes in the communication characteristics and patterns of the devices. Such changes can be caused by the attack itself or be consequences of a successful attack. The considered attacks can be roughly divided into two categories:
\begin{itemize}
\item Attacks in which the attacker attempts to compromise, remotely control, or render a device unusable through the network. Examples include DoS attacks, scans, unauthorized remote access, and the sending of control messages that were not initiated by the Smart Home resident.
\item Attacks in which the attacker misuses a compromised device to perform the above attacks against other devices or to exfiltrate data.
\end{itemize}
We assume that the attacker has the capabilities to attempt the above attacks from inside and outside the Smart Home network. Obviously, our approach cannot protect against attacks that do not affect the network communications that are visible to the firewall. Those include attacks with direct physical access to a device (e.g., stealing the SD card of an IP camera, physically manipulating a door sensor, etc.), and attacks where the network communication stays inside the limits and patterns described in the profiles.

We assume that the attacker does not have the capability to
\begin{enumerate}
\item compromise the firewall software or the equipment hosting the firewall;
\item compromise the device profiles.
\end{enumerate}
The second point deserves further explanation. In the MUD standard \cite{mud}, the idea is that the MUD file is provided by the manufacturer. It can be debated whether end users should trust the manufacturer when it comes to the protection of their Smart Homes. Manufacturers have been caught collecting sensitive information from their customers' devices on several occasions \cite{zavalyshyn2020}. Therefore, the possibility that a manufacturer distributes compromised profiles cannot be excluded. We have designed our profile description language such that the resulting profiles are easy for experts to understand. We envisage a trusted platform from which profiles can be securely downloaded and where experienced users can review them. Since this approach has already been discussed for other use cases in the literature \cite{zavalyshyn2020, guo_realizing_2023}, we consider the design and implementation of such a platform to be outside the scope of this paper.

\section{Overview and design principles}
\label{sec:overview}

In this section, we give a high-level introduction to our security system for Smart Home networks. We provide an overview on its components and how they work together to implement our smart firewall. We also describe the threat model that our system is designed to protect against.

\subsection{System overview}

Our approach is based on extended traffic profiles that describe the allowed communication patterns, including device interactions, of Smart Home devices. Our goal is to defend against attacks by using a firewall that leverages the information contained in the traffic profiles to identify and block unwanted or unexpected communications. As explained in Section~\ref{sec:smart-home}, we assume in this paper that the Smart Home to be protected is based on a (wireless) LAN infrastructure. An obvious place for a firewall is therefore the central home router or access point, where the communication between the devices and with the outside world can be monitored. If other communication channels exist for the devices (e.g., ad-hoc or non-WiFi communication), they have to be monitored, too, as discussed in Sec.~\ref{sec:discussion}.

\begin{figure}
    \centering

    \tikzstyle{traffic} = [->, draw=blue]
    \begin{tikzpicture}[shorten >=1pt,semithick,>={stealth},node distance=1cm,on grid,auto]

        
        \node (profile-1) {\includegraphics[scale=0.04]{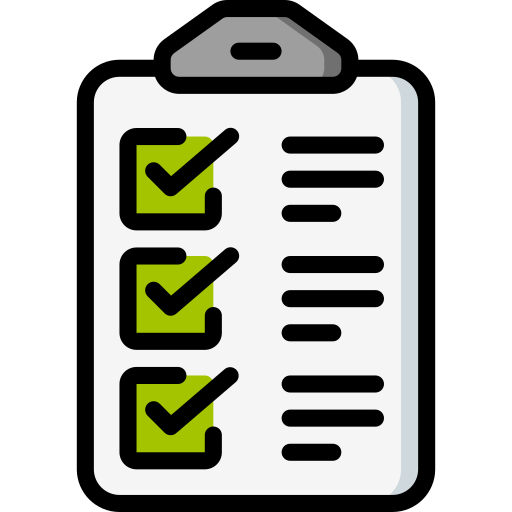}};
        \node (profile-2) [above of=profile-1] {\includegraphics[scale=0.04]{images/intro/clipboard.png}};
        \node (profile-3) [label={below:\scriptsize Profiles}, below of=profile-1] {\includegraphics[scale=0.04]{images/intro/clipboard.png}};
        \node (translator) [label={[yshift=-0.2cm]above:Translator}, right=1.6cm of profile-1] {\includegraphics[scale=0.05]{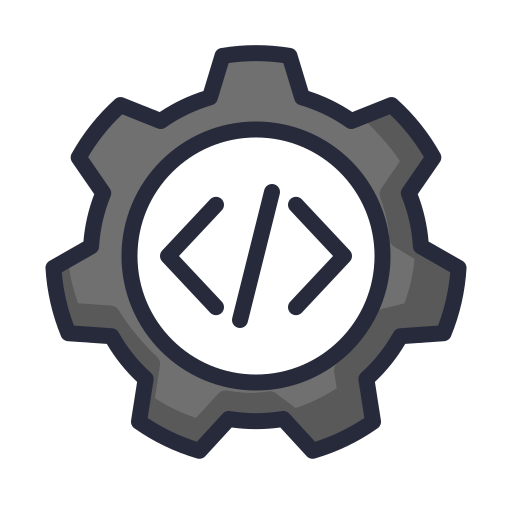}};
        \node (template) [label={[yshift=0.1cm]below:Templates}, below=1.3cm of translator] {\includegraphics[scale=0.04]{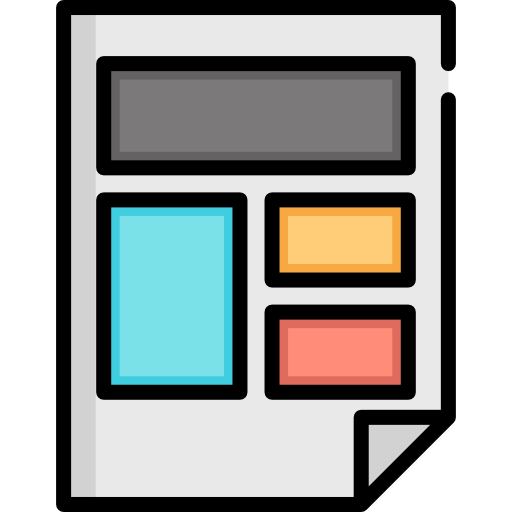}};
        \node (nft-script) [label={[yshift=-0.2cm,align=center]above:NFTables\\script}, above=0.7cm of translator, xshift=1.6cm] {\includegraphics[scale=0.05]{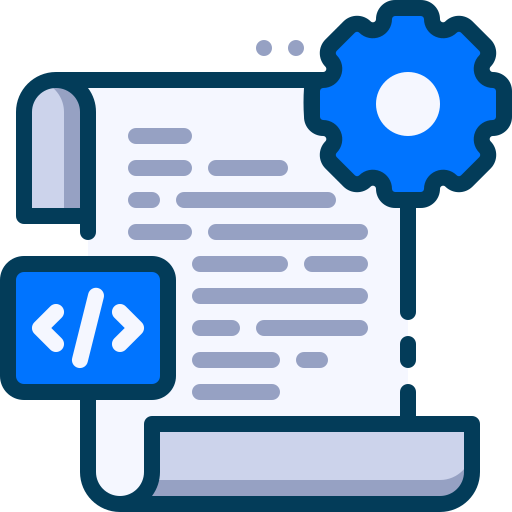}};
        \node (nfq-src) [label={[yshift=0.15cm,align=center]below:NFQueue\\C program}, below=0.7cm of translator, xshift=1.6cm] {\includegraphics[scale=0.05]{images/system-overview/script.png}};

        \node (nft-firewall) [label={[yshift=-0.2cm,align=center]above:NFTables\\firewall}, right=1.5cm of nft-script] {\includegraphics[scale=0.05]{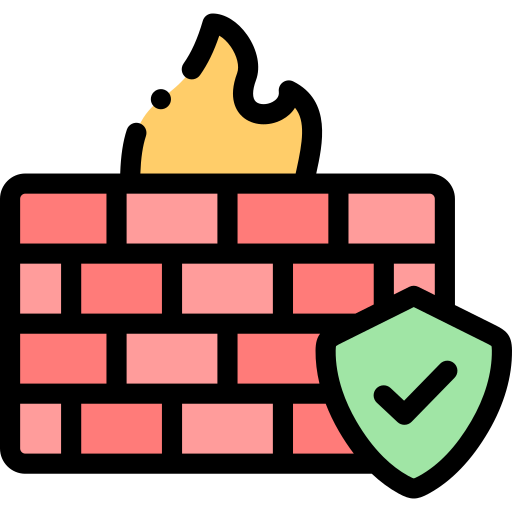}};
        \node (nfq-exec) [label={[yshift=0.15cm,align=center]below:NFQueue\\executable}, right=1.5cm of nfq-src] {\includegraphics[scale=0.05]{images/system-overview/development.png}};
        \node (router) [label={[yshift=0.15cm,align=center]below:Router}, right=1.7cm of nft-firewall, yshift=-0.7cm] {\includegraphics[scale=0.05]{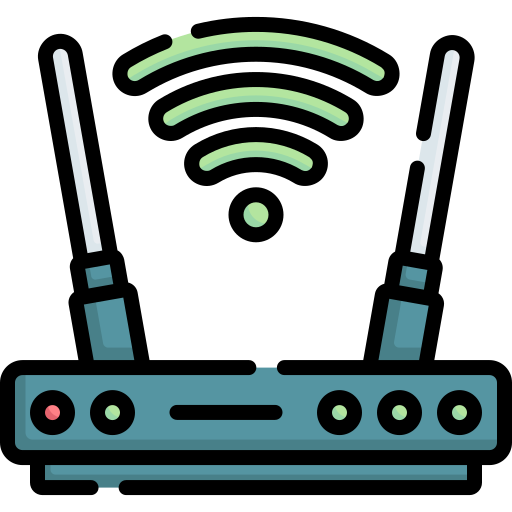}};

        \node (device-1) [label={[align=center, xshift=0.5cm]above:Devices}, above=1.3cm of router, xshift=-0.5cm] {\includegraphics[scale=0.04]{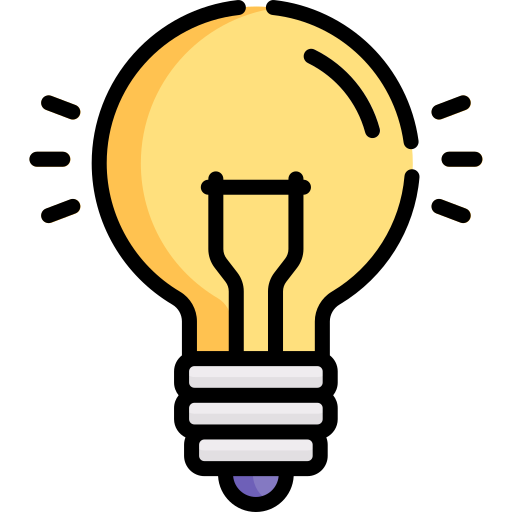}};
        \node (device-2) [above=1.3cm of router, xshift=0.5cm] {\includegraphics[scale=0.03]{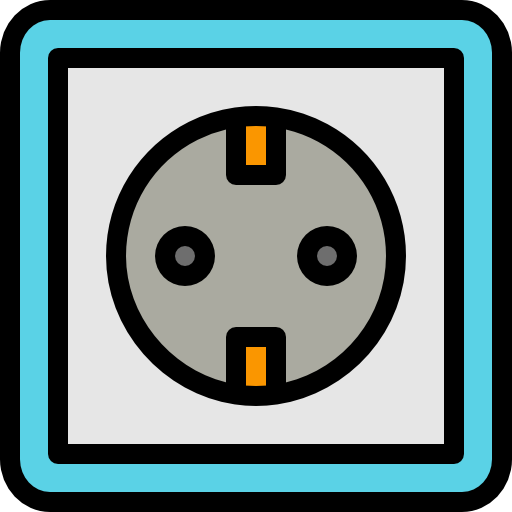}};
        \node (packet) [above=0.8cm of router] {\includegraphics[scale=0.025]{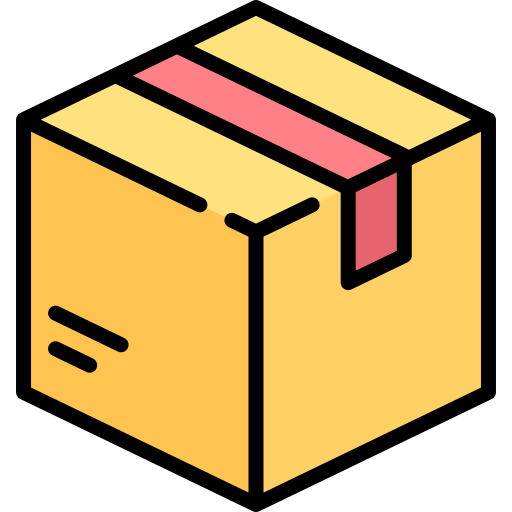}};
        \node (accept) [below=0.7cm of router.south, xshift=-0.25cm] {\includegraphics[scale=0.021]{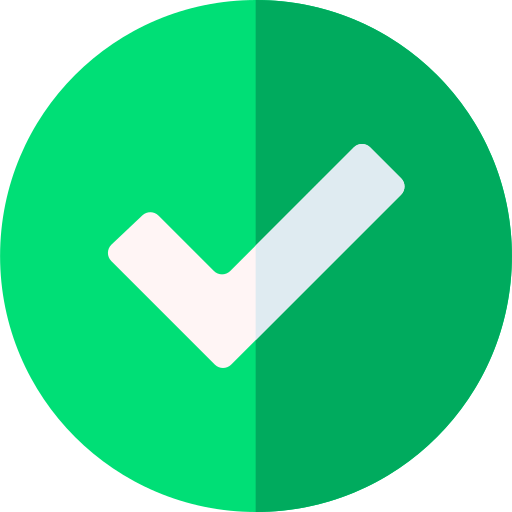}};
        \node (drop) [below=0.7cm of router.south, xshift=0.25cm] {\includegraphics[scale=0.02]{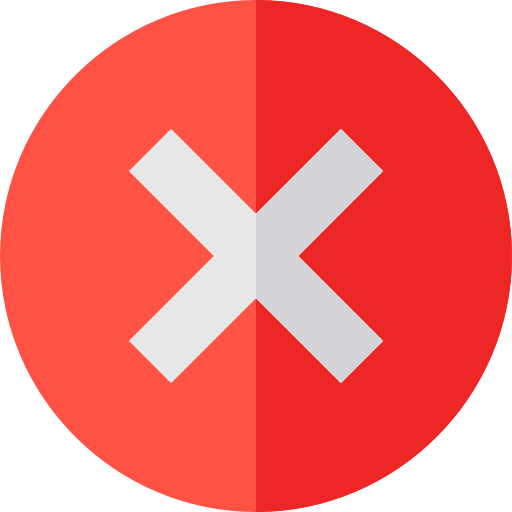}};

        \node[circle, fill=black, inner sep=1pt] (profiles-merge) [right=0.7cm of profile-1] {};
        \node (translator-split) [right=0.8cm of translator] {};
        \node[circle, fill=black, inner sep=1pt] (firewall-merge) [left of=router] {};


        \draw [->] (profile-1) -- (translator);
        \draw      (profile-2) -| ([yshift=-0.04cm]profiles-merge.center);
        \draw      (profile-3) -| ([yshift=0.04cm]profiles-merge.center);
        \draw [->] (template) -- (translator);

        \draw (translator) -- (translator-split.center);
        \draw [->] ([xshift=-0.040cm]translator-split.center) |- (nft-script);
        \draw [->] ([xshift=-0.040cm]translator-split.center) |- (nfq-src);

        \draw [->] (nft-script) -- node[above] {} (nft-firewall);
        \draw [->] (nfq-src)    -- node[above] {} (nfq-exec);
        
        \draw (nft-firewall) -| ([yshift=-0.04cm]firewall-merge.center);
        \draw (nfq-exec) -| ([yshift=0.04cm]firewall-merge.center);
        \draw [->] (firewall-merge.center) -- (router);

        \draw [traffic] (device-1) -- (router);
        \draw [traffic] (device-2) -- (router);
        \draw [traffic] ([yshift=-0.2cm]router.south) -- (accept);
        \draw [traffic] ([yshift=-0.2cm]router.south) -- (drop);


        \begin{scope}[on background layer]

            \node (compile-bottom-left) [below of=profile-3, xshift=-0.6cm]  {};
            \node (compile-top-right)   [above of=nft-script, xshift=0.7cm] {};
            \fill[rounded corners, blue!10] (compile-bottom-left) rectangle (compile-top-right) {};
            \node (compile-legend) [below=0.3cm of compile-bottom-left, xshift=2.3cm] {\scriptsize \textbf{Setup time}};

            \node (run-bottom-left) [below=1.3cm of nfq-exec, xshift=-0.7cm] {};
            \node (run-top-right)   [above=2.1cm of router, xshift=1.1cm] {};
            \fill[rounded corners, orange!10] (run-bottom-left) rectangle (run-top-right) {};
            \node (run-legend) [below=0.3cm of run-bottom-left, xshift=1.8cm] {\scriptsize \textbf{Run time}};

        \end{scope}

    \end{tikzpicture}

    \caption{System components}
    \label{fig:system}
\end{figure}

Figure~\ref{fig:system} shows the components of our proposed system. On the left we have the device profiles, one for each device in the Smart Home network. Profiles are written in a human-readable simple language by the manufacturer of the devices, by the device owner, or other users. They describe the allowed communication activities and their characteristics at different protocol and abstraction levels. Those comprise connection parameters (endpoints, protocols, etc.), traffic statistics (number of packets, byte rate, etc.), application layer information (e.g., URIs), and the relationships between connections (such as the sequence, for example). 

Our firewall, running on a router in Figure~\ref{fig:system}, builds on NFTables \cite{nftables}, the novel Linux kernel firewall, to filter traffic from and to the Smart Home devices. It installs NFTables rules that, where necessary, offload matching packets to a user-space program using the NFQueue infrastructure. This program is responsible for performing the extended matching and filter actions specified in the profiles.
The operation of NFTables and NFQueue will be described in more detail in Section~\ref{sec:firewall}.

Our firewall is intended for use on consumer-grade home routers and access points. For efficiency reasons, it does not load the profiles during runtime. Instead, a translator (also written by us) parses the profiles and emits, guided by a set of code templates, NFTables configuration scripts and the code for the user-space program.

\section{System Description}
\label{sec:system}

In the following, we describe the system shown in Figure~\ref{fig:system} in detail. We first describe the structure and language of the device profiles and then explain how the firewall uses them to block unwanted traffic.

\subsection{Devices profiles}
\label{sec:profiles}

A device's profile is a model of its legitimate network communication patterns.
We use YAML \cite{yaml} as file format, instead of the JSON or XML formats used by MUD. The main reason for this choice is that the YAML syntax is more flexible, allowing, for example, the use of references to other fields in the same file. In the case of repeating patterns, this improves readability and reduces cluttering. We extend this concept to allow referencing fields present in \emph{other} files and also placeholders. To this end, our profile language supports a new \texttt{include} directive, which we have implemented thanks to \texttt{PyYAML}'s support for custom YAML parsers in Python.

Nevertheless, the features supported by our profiles are a superset of MUD, i.e., any MUD file can be trivially translated into our profile format.

\begin{figure}
    \centering
    \includegraphics[width=\linewidth]{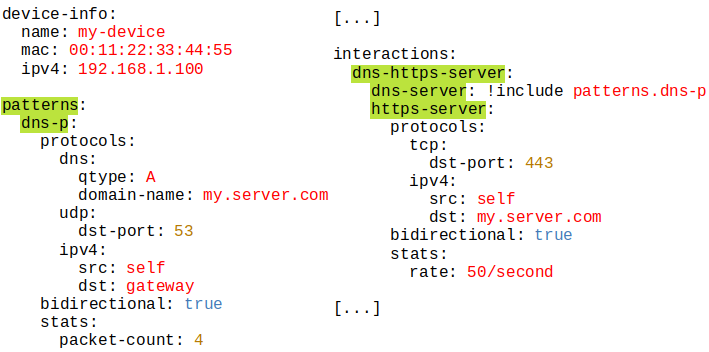}
    \caption{
        Minimal device profile example.
        User defined labels are highlighted in green. Builtin keywords are shown in black.
    }
    \label{fig:profile}
\end{figure}

Figure~\ref{fig:profile} shows an example of a small (and incomplete) device profile. It consists of two main sections, namely the \texttt{device-info} section and the \texttt{interactions} section, that we describe in the following. The example also showcases the \texttt{include} directive that allows to move parts of a declaration to another location in the profile and to refer to it by a qualified name.
Concretely, the identifier \texttt{patterns.dns-p} refers to the field \texttt{dns-p} under the field \texttt{patterns}. Both field names are user defined and highlighted in green in the figure.
A larger device profile is showcased on Figure~\ref{fig:profile-full}, in Appendix.

\subsubsection{Device metadata}

The \texttt{device-info} field specifies metadata used to identify the device, that is its name, MAC address, and IP address (v4 or v6). In the remaining of the profile, the keyword \texttt{self} can be used to reference the device.

\subsubsection{Interactions}
\label{sec:interactions}

The \texttt{interactions} section defines a series of \emph{interactions}. In the example, we only have one, which we (the profile author) have named \texttt{dns-https-server}. An interaction defines a series of \emph{policies}. In our example, the \texttt{dns-https-server} interaction defines two policies that we have named \texttt{dns-server} and \texttt{https-server}. We will first discuss them before explaining what interactions do.

The firewall only lets through packets matching a policy. The latter describes the packet features to match on. We allow all features supported by MUD (e.g., IP addresses, TCP/UDP ports, ARP message types, ICMP message types). In addition, we added support for more protocols, in particular for the application layer protocols (m)DNS, DHCP, HTTP, SSDP, CoAP, and IGMP, as these protocols are used by many Smart Home devices.

To describe more complex network activities, multiple policies can be combined into an interaction definition (\texttt{dns-https-server} in the example). In an interaction, not all its policies are active at the same time.
An interaction starts when its first policy sees a matching packet. That first policy becomes then the active policy of the interaction. Following rules that we explain below, the interaction will at some point deactivate its currently active policy and activate its next policy. How this is done depends on the type of the currently active policy:

By default, a policy is a \emph{one-off} policy. When it matches a packet, it is deactivated and the next policy is activated. If the property \texttt{bidirectional} is set, the policy requires a matching packet in each direction: it first matches on the fields as specified and then inverses the relevant fields (source and destination addresses and ports, DNS or ICMP message types, etc.) for the next packet.

A \emph{transient} policy has a maximum duration or maximum packet count specified in its \texttt{stats} section. The interaction moves to the next policy if the transient policy has reached the specified maximum duration or number of matching packets \emph{or} when a packet is seen that matches the next policy. If the property \texttt{bidirectional} is set, the policy matches packets in both directions as long as it is active.

A \emph{periodic} policy contains a packet rate specification in the \texttt{stats} section. The policy stays active until a packet is seen that matches the next policy. Packets exceeding the specified rate are dropped.

\subsection{The Firewall}
\label{sec:firewall}

In the following, we describe a firewall that is able to enforce the traffic patterns described in our profile language.
It leverages NFTables, the next-generation Linux kernel firewall \cite{nftables} and successor of IPTables \cite{purdy2004linux}. Since NFTables is shipped by default in many recent Linux distributions, our firewall can run on many platforms and does not require installing third-party tools.

\subsubsection{A quick introduction to NFTables}

NFTables provides simple rule-based packet header matching for the most prominent layer 2, 3 and 4 protocols, such as ARP, IPv4 and IPv6, TCP, UDP and ICMP, as well as matching capabilities based on rate and size. NFTables allows intercepting packets traversing the Linux network stack in multiple location using the \emph{Netfilter kernel hooks} \cite{netfilter-hooks}. A packet passing through a hook will trigger the callback execution of an attached function. In our case, to ensure early traffic filtering, our firewall registers NFTables to the \texttt{prerouting} hook, which is the first hook able to detect packets coming from different interfaces. For further explanations on the different hook types, we refer to the extensive documentation of Netfilter.

\begin{figure}
    \centering

    \tikzstyle{accept} = [->, draw=ForestGreen]
    \tikzstyle{drop}   = [->, draw=red]
    \tikzstyle{queue}  = [->, draw=blue]
    \begin{tikzpicture}[shorten >=1pt,semithick,>={stealth},node distance=2cm,on grid,auto]


        \node (pkt)  [label={[yshift=0.15cm]below:Packet}]  {\includegraphics[scale=0.04]{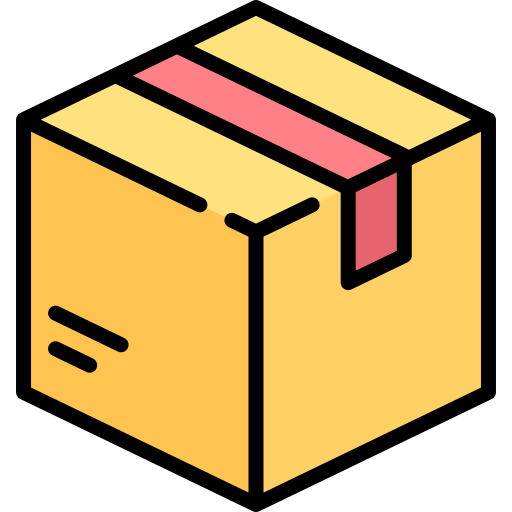}};
        \node (hook) [label={[align=center, yshift=0.15cm]below:\texttt{prerouting}\\hook}, right=1.5cm of pkt] {\includegraphics[scale=0.03]{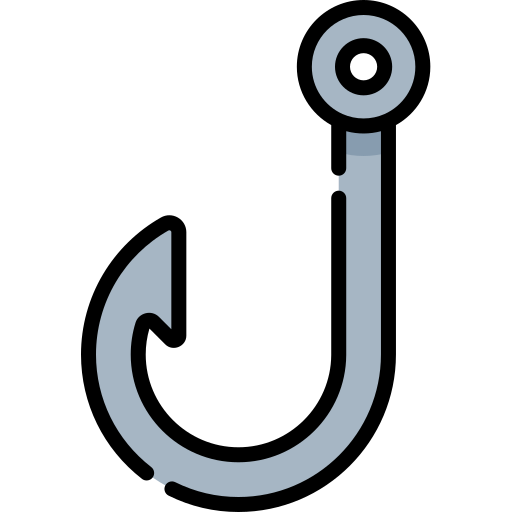}};
        \node (nft)  [label={[yshift=0.15cm]below:NFTables}, right=1.5cm of hook] {\includegraphics[scale=0.05]{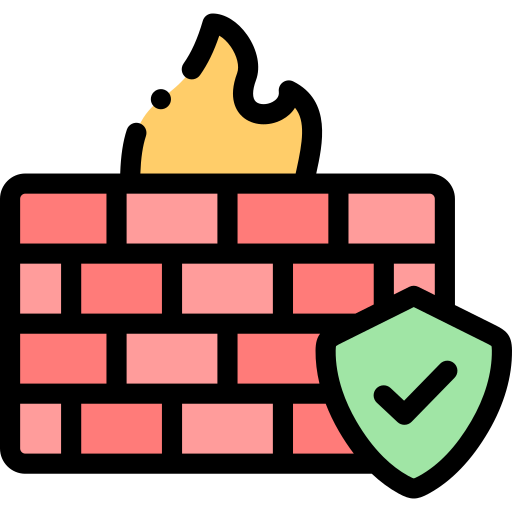}};
        \node (nft-rules) [above=0.7cm of nft] {\includegraphics[scale=0.04]{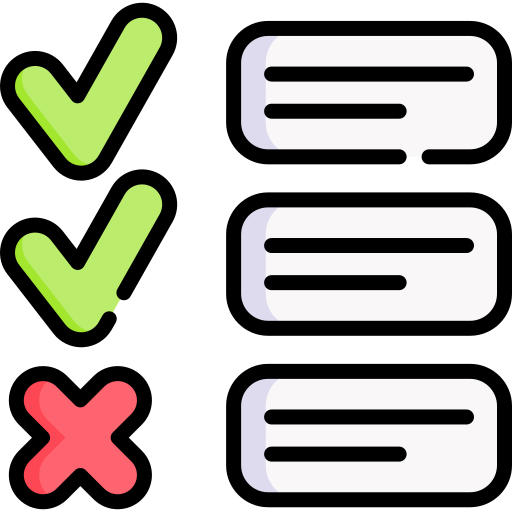}};
        \node (nfq)  [label={[yshift=0.15cm]below:NFQueue}, right=1.7cm of nft] {\includegraphics[scale=0.05]{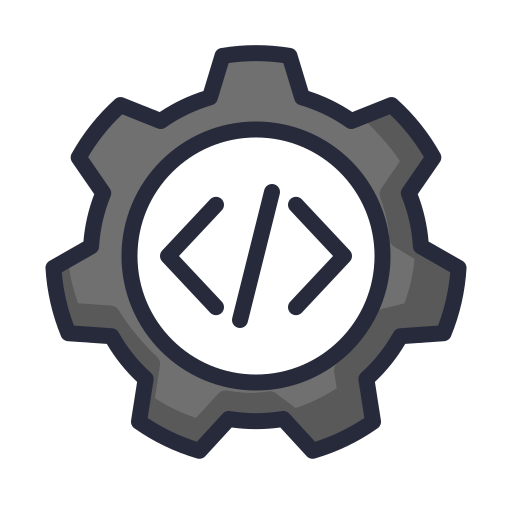}};
        \node (nfq-src) [above=0.7cm of nfq] {\includegraphics[scale=0.04]{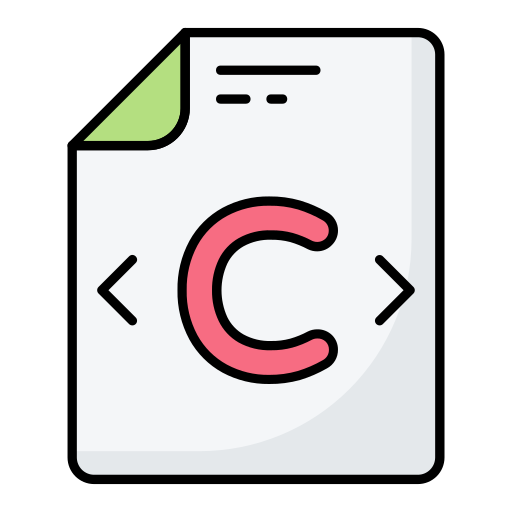}};
        
        \node (accept-nft) [below=1.3cm of nft,xshift=-0.3cm]  {\includegraphics[scale=0.025]{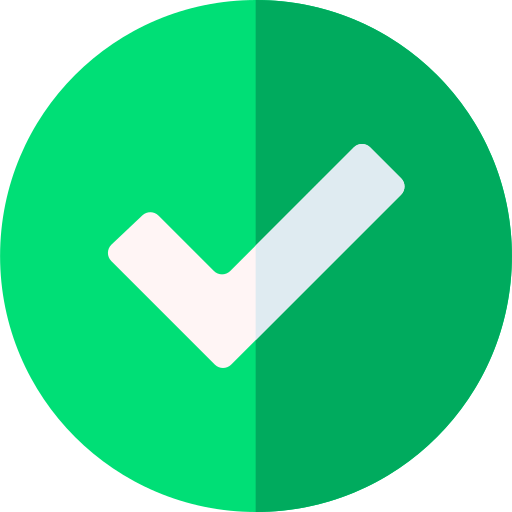}};
        \node (drop-nft)   [below=1.3cm of nft,xshift=0.3cm] {\includegraphics[scale=0.025]{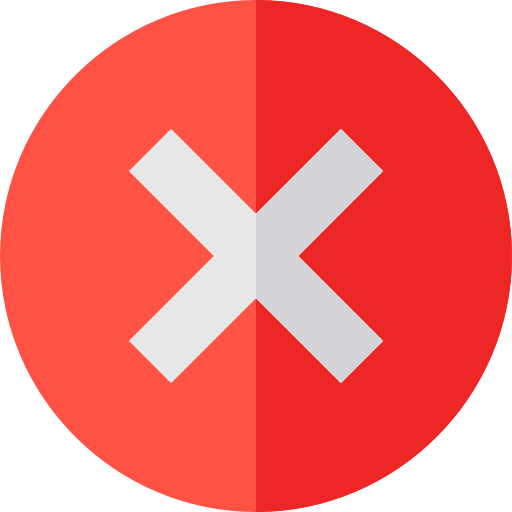}};
        \node (accept-nfq) [below=1.3cm of nfq,xshift=-0.3cm]  {\includegraphics[scale=0.025]{images/firewall/check.png}};
        \node (drop-nfq)   [below=1.3cm of nfq,xshift=0.3cm] {\includegraphics[scale=0.025]{images/firewall/cancel.png}};

        \begin{scope}[on background layer]

            \node (user-bottom-left) [below=1.7cm of nfq, xshift=-0.8cm]  {};
            \node (user-top-right)   [above=1.1cm of nfq, xshift=0.8cm] {};
            \fill[rounded corners, blue!20] (user-bottom-left) rectangle (user-top-right) {};
            \node (user-legend) [below=0.01cm of user-bottom-left, anchor=north west] {\textbf{User-space}};

            \node (kernel-bottom-left) [below=1.7cm of hook, xshift=-0.7cm]  {};
            \node (kernel-top-right)   [above=1.1cm of nft, xshift=0.7cm] {};
            \fill[rounded corners, orange!30] (kernel-bottom-left) rectangle (kernel-top-right) {};
            \node (kernel-legend) [below=0.01cm of kernel-bottom-left, anchor=north west] {\textbf{Kernel}};

            \node (stack-bottom-left) [below=0.85cm of hook, xshift=-0.6cm]  {};
            \node (stack-top-right)   [above=1cm of hook, xshift=0.6cm] {};
            \fill[rounded corners, orange!60] (stack-bottom-left) rectangle (stack-top-right) {};
            \node (stack-legend) [left=1.2cm of stack-top-right, anchor=north west, text width=1cm] {\textbf{Network stack}};

        \end{scope}

        \draw [->] (pkt) -- (hook);
        \draw [->] (hook) -- (nft);
        \draw [queue] (nft) -- (nfq);
        \draw [accept] ([yshift=-0.2cm]nft.south) -- ([yshift=0.2cm]accept-nft.center);
        \draw [drop] ([yshift=-0.2cm]nft.south) -- ([yshift=0.2cm]drop-nft.center);
        \draw [accept] ([yshift=-0.2cm]nfq.south) -- ([yshift=0.2cm]accept-nfq.center);
        \draw [drop] ([yshift=-0.2cm]nfq.south) -- ([yshift=0.2cm]drop-nfq.center);
    \end{tikzpicture}

    \caption{General NFTables/NFQueue firewall operation}
    \label{fig:firewall}
\end{figure}

By itself, NFTables' matching and filtering capabilities are quite limited. Its flexibility comes from one of its libraries, \texttt{libnetfilter\_queue} (hereafter referred to as NFQueue),
which allows NFTables rules to offload matching packets to a so-called ``queue'',
implemented as an arbitrary callback function in a user-space program for further treatment \cite{nfqueue}. The latter can, in turn, process the packet and issue the final verdict (i.e., accept or drop), as summarized in Figure~\ref{fig:firewall}.

\subsubsection{User-space packet queuing with NFQueue}
\label{sec:nfqueue}

We use NFTables' limited packet header matching capabilities to execute the basic matching rules defined in the policies of our profiles, e.g., the matching by transport protocol or IP address. We implement the more advanced features of our firewall in external binaries that receives the packets from NFTables using NFQueue. Those features are:
\begin{enumerate}
\item Support for the protocols CoAP, DHCP, HTTP, IGMP, (m)DNS, and SSDP.
\item Support for domain names. Our firewall monitors DNS responses and caches the resolved names in an internal table. This allows using domain names in policies instead of IP addresses, which is necessary for our profiles to stay up-to-date,
as the IP addresses of the cloud servers contacted by an IoT device might regularly change.
\item Matching on maximum duration and packet count, as needed for transient policies.
\item The interaction mechanism.
\end{enumerate}

Our system currently does not support matching on encrypted data. First of all, it is not always possible to decrypt traffic. Second, the firewall might be deployed on a small consumer grade device that does not have the necessary resources to decrypt all traffic.
It should be noted, however, that many IoT devices use clear text protocols due to resource constraints or because the manufacturer is shy of the associated effort (updating certificates, etc.). Furthermore, previous research has shown that inspecting only the headers and metadata of network traffic can provide sufficient information for detecting many types of malicious traffic \cite{nobakht_host-based_2016, pashamokhtari_progressive_2020}.

\subsubsection{State machine for interactions}
\label{sec:fsm}

The interaction-based matching and filtering mechanism described in Section~\ref{sec:interactions} requires a stateful firewall. Our current prototype maintains a Finite State Machine (FSM) per interaction, where the states correspond to the policies of the interaction.
The number of states further depends on the types of the policies, on whether they are bidirectional.

\begin{figure*}
    \centering
  
    \tikzset{
        every node/.style={font=\footnotesize},
        every state/.append style={minimum size=10pt}
    }
    \definecolor{fsm-green}{cmyk}{1,0,1,0} 
    \begin{tikzpicture}[shorten >=1pt,semithick,>={stealth},node distance=3.5cm,on grid,auto]
        
        \node[state,initial]  (s_0)                {0};
        \node[state]          (s_1) [right=of s_0] {1};
        \node[state]          (s_2) [right=of s_1] {2};
        \node[state]          (s_3) [right=of s_2] {3};
        
        \path[->] (s_0) edge [loop above] node {*: \textcolor{red}{drop}} ()
                        edge              node {$A$, fwd: \textcolor{fsm-green}{accept}} (s_1)
                    (s_1) edge [loop above] node {*: \textcolor{red}{drop}} ()
                        edge              node {$A$, bwd: \textcolor{fsm-green}{accept}} (s_2)
                    (s_2) edge [loop above] node [align=center] {*: \textcolor{red}{drop}; $B$: \textcolor{fsm-green}{accept}} ()
                        edge              node {$C$: \textcolor{fsm-green}{accept}} (s_3)
                    (s_3) edge [loop above] node [align=center] {*: \textcolor{red}{drop}; $C$, below: \textcolor{fsm-green}{accept}} ()
                        edge [bend left=25, below] node [above] {$C$, above: \textcolor{fsm-green}{accept}} (s_0)
                        edge [bend left=22, below] node [above] {$A$, fwd: \textcolor{fsm-green}{accept}} (s_1);
        \end{tikzpicture}
\vspace{-3mm}
    \caption{
        Finite State Machine representing an interaction consisting of a one-off policy $A$, a periodic policy $B$ and a transient policy $C$. ``fwd'' and ``bwd'' specify the direction of the packet. ``below'' indicates a packet within the limits of a transient policy. ``*'' stands for all non-matching traffic.
    }
    \    \label{fig:fsm}
\end{figure*}
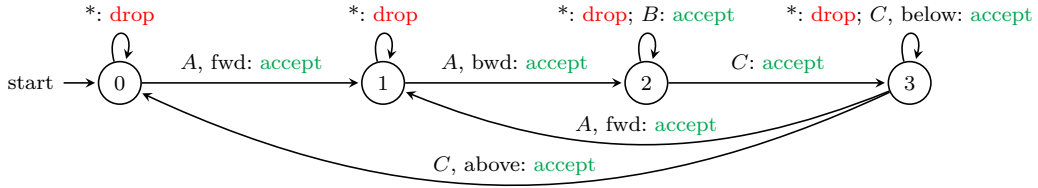

We explain the functioning of the FSM with an example of an interaction with three policies $A$, $B$, and $C$. Policy $A$ is one-off and bidirectional, $B$ is periodic, and $C$ is transient. The resulting FSM is shown in Figure~\ref{fig:fsm}.

At the beginning, the FSM is in the initial state 0. If a packet matching policy $A$ is seen, the FSM moves to state 1. Since $A$ is bidirectional, the FSM now waits for the packet in the opposite direction. After that, it moves to state 2, which presents policy $B$. That policy is periodic, i.e., it will accept all matching packets and stay in state 2 until a packet matching policy $C$ is seen, which causes the FSM to move to state 3. As a transient policy, $C$ accepts all matching packets, but, unlike policy $B$, it will stop matching when the specified maximum duration or packet count is reached. In that case, the interaction restarts (transition to state 0) or the FSM directly moves to state 1 if the last packet matched policy $A$. All non-matching packets are dropped. Note that the rate limitation of the periodic policy is directly enforced by NFTables rules and not by the FSM.

\subsection{The Translator}
\label{sec:translator}

The translator is a tool written by us in Python that translates the device profiles into NFTables configuration scripts and C code for the NFQueue part of the firewall. To simplify the code generation, we have prepared code templates with placeholders that are filled with the information extracted from the profiles using the Jinja templating engine \cite{jinja}. The C code is then compiled to binary files that are finally executed on the device hosting the firewall.
When the firewall is running, there is one user-space process per device, which corresponds to the device's NFQueue executable. 
This execution is further split into multiple threads, one per queue, that is, one per NFTable rule implying the device.
A packet involving multiple devices triggers all respective NFQueue executables. It is admitted if it complies to all prompted policies.

\section{Experimental setup}
\label{sec:setup}

This section describes the experimental setup that we use for the evaluation of our approach and the prototype implementation of the firewall. It represents a typical home environment,
composed of commercial, off-the-shelf devices.

\subsection{Devices}
\label{sec:testbed-devices}

\begin{table*}
    \centering
    \small
    \begin{tabular}{|c|c|c|c|c|c|c|c|}
    \hline
    \textbf{ID} & \textbf{Name} & \textbf{Type} & \textbf{Protocol} & \textbf{Ref} & \textbf{\#Policies} & \textbf{\#Inter.} & \textbf{\#Queues}\\
    \hline
    1 & TP-Link HS110 smart plug & End device & Wi-Fi & \cite{hs110} & 35 & 13 & 20 \\
    2 & Xiaomi MJSXJ02CM camera & End device & Wi-Fi & \cite{xiaomi-cam} & 37 & 15 & 21 \\
    3 & D-Link DCS-8000LH camera & End device & Wi-Fi & \cite{dlink-cam} & 58 & 18 & 24 \\
    4 & Philips Hue color lamp & End device & Zigbee & \cite{color-lamp} & / & / & / \\
    5 & SmartThings door sensor & End device & Zigbee & \cite{multi-sensor} & / & / & / \\
    6 & SmartThings plug & End device & Zigbee & \cite{st-plug} & / & / & / \\
    7 & SmartThings presence sensor & End device & Zigbee & \cite{presence-sensor} & / & / & / \\
    8 & Amazon Echo Dot (2nd gen) & Hub & Wi-Fi & \cite{echo-dot} & / & / & / \\
    9 & Philips Hue bridge & Gateway & Ethernet, Zigbee & \cite{hue-bridge} & 96 & 33 & 26 \\
		10 & SmartThings V3 hub & Hub & Ethernet, Zigbee & \cite{smartthings-hub} & 41 & 18 & 17  \\
    \hline
    \end{tabular}
    \caption{
      Testbed devices. The ID column is used to identify each device in Figure~\ref{fig:network}.\\
      Devices 4, 5 \& 6 do not have associated rules because they use the Zigbee protocol.\\
      Device 8 does not have associated rules as it was only use as a hub.\\
      (Inter. = Interactions)
    }
    \label{tab:devices}
\end{table*}

To mimic a typical Smart Home network, we instrument a set of commercial, off-the-shelf devices (Table~\ref{tab:devices}).
Smart Home devices can be divided into two groups, namely \emph{end devices} and \emph{Smart Hubs}. The former include sensors and actuators, i.e., devices that sense and manipulate the physical environment in a Smart Home.
Some end devices use the Zigbee protocol \cite{ergen2004zigbee}, a protocol designed for low-rate communication with resource and energy constrained IoT devices, and require an intermediate device (a \emph{gateway}) to communicate with IP networks. Smart Hubs generally are more complex devices that allow to control other devices and have built-in additional functions such as a Smart Speaker. Many commercially available hubs have multiple network interfaces for different types of networks and can therefore also act as a gateway. 

We connect the IP devices, either via Wi-Fi or Ethernet, to a router acting as a wireless Access Point (AP), running OpenWrt \cite{openwrt}, an open-source Linux distribution specifically designed for network appliances.
The AP is in turn wired to a router, also running OpenWrt, which acts as the gateway between the home LAN and the Internet. Our firewall is deployed on the AP, which is a very modest dual-band AP (TP-Link TL-WDR4900, v1) with a 800~MHz Freescale PPC CPU and 128~MBytes of RAM.

\begin{figure}
  \centering
  
  \begin{tikzpicture}[node distance=2cm,on grid,auto]

    
      \node (ap)[label={[label distance=-0.1cm]above:AP}]{\includegraphics[scale=0.06]{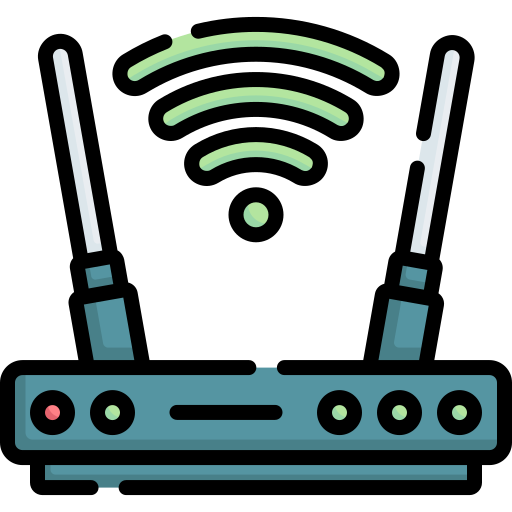}};
      \node (router)[right=1.5cm of ap,label={[label distance=-0.25cm]above:Router}]{\includegraphics[scale=0.07]{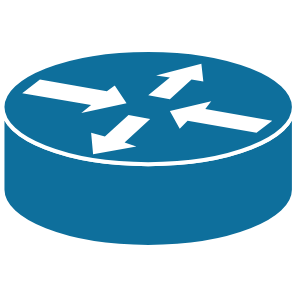}};

      \node (phone) [above=1cm of ap,xshift=-1.5cm] {\includegraphics[scale=0.05]{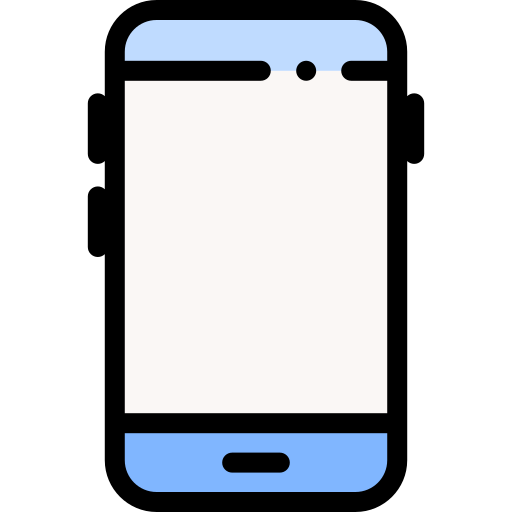}};
      \node (wi-fi-plug)[left=2.1cm of ap,label={[label distance=-0.09cm]left:1}]{\includegraphics[scale=0.04]{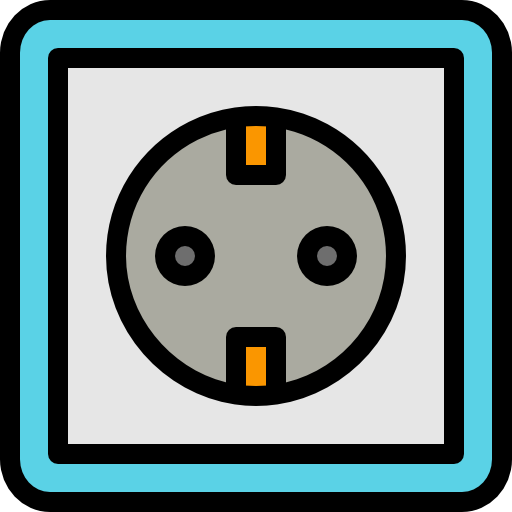}};
      \node (camera-1)[left=2.1cm of ap,yshift=-0.8cm,label={[label distance=-0.09cm]left:2}]{\includegraphics[scale=0.05]{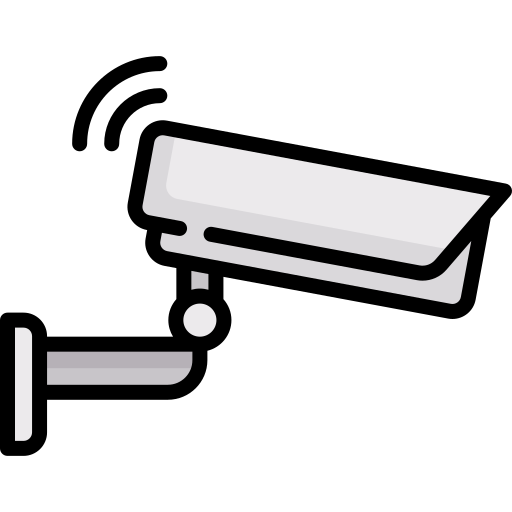}};
      \node (camera-2)[left=2.1cm of ap,yshift=-1.6cm,label={[label distance=-0.09cm]left:3}]{\includegraphics[scale=0.05]{images/setup/camera.png}};
      \node (alexa)[left=2.1cm of ap,yshift=-2.4cm,label={[label distance=-0.15cm]left:8}]{\includegraphics[scale=0.05]{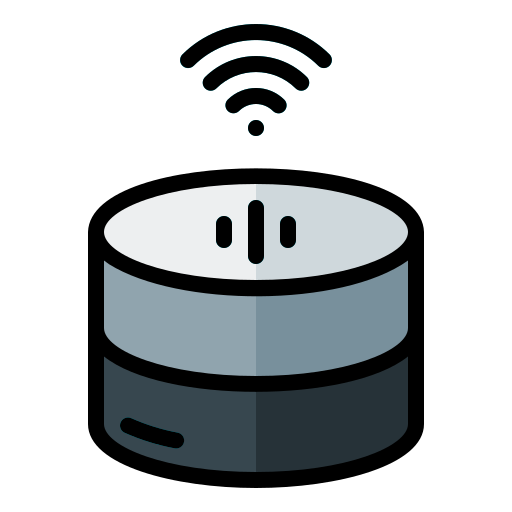}};

      \node (hue-bridge)[below=1.3cm of ap,xshift=-0.4cm,label={[label distance=-0.15cm]right:9}]{\includegraphics[scale=0.05]{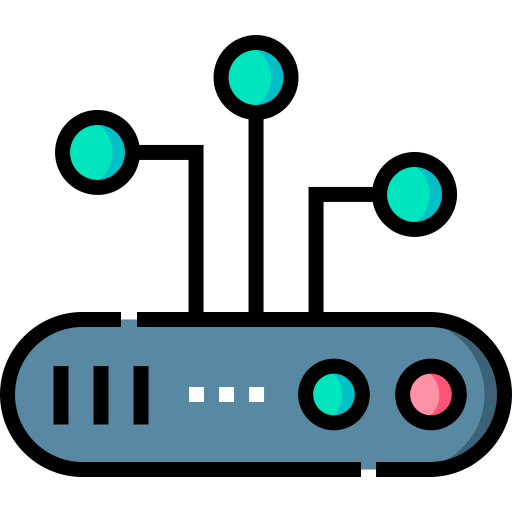}};
      \node (st-hub)[below=1.3cm of ap,xshift=0.8cm,label={[label distance=-0.15cm]right:10}]{\includegraphics[scale=0.05]{images/setup/gateway.png}};

      \node (light-bulb)[below=1.2cm of hue-bridge,xshift=-0.6cm,label={[label distance=-0.09cm]below:4}]{\includegraphics[scale=0.05]{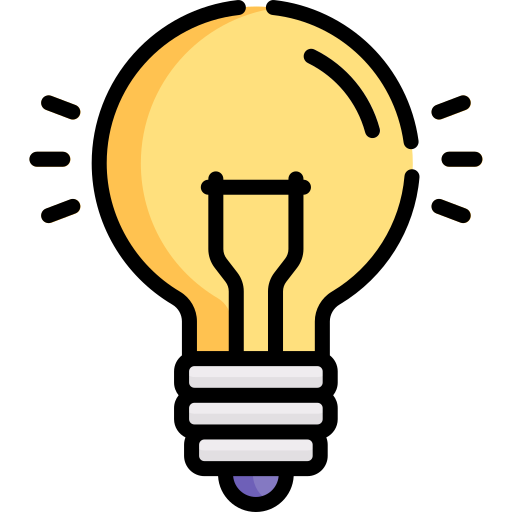}};
      \node (door-sensor)[below=1.2cm of st-hub,xshift=-0.9cm,label={[label distance=-0.09cm]below:5}]{\includegraphics[scale=0.05]{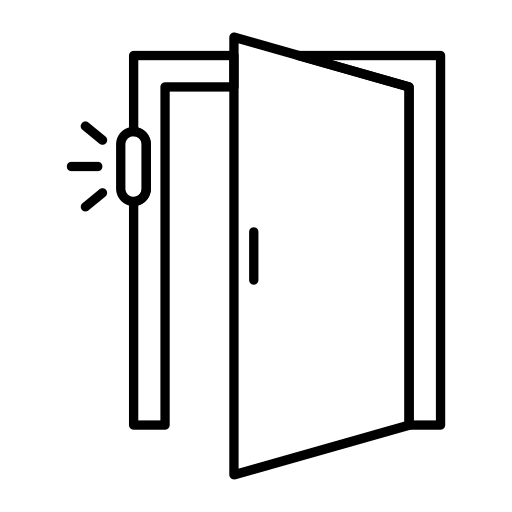}};
      \node (st-plug)[below=1.2cm of st-hub,label={[label distance=-0.09cm]below:6}]{\includegraphics[scale=0.04]{images/setup/power-socket.png}};
      \node (presence-sensor)[below=1.2cm of st-hub,xshift=0.9cm,label={[label distance=-0.09cm]below:7}]{\includegraphics[scale=0.04]{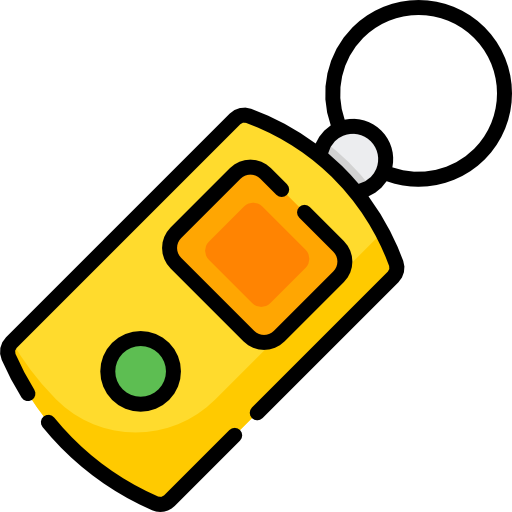}};

      \node (internet)[right=1.7cm of router]{\includegraphics[scale=0.08]{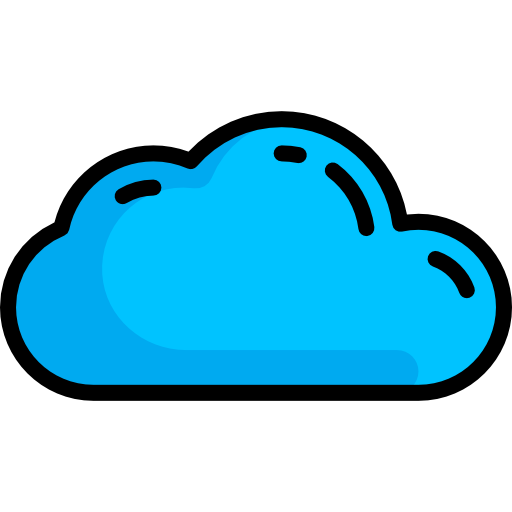}};


      \draw [wireless] (phone)       -- (ap);
      \draw [wireless] (wi-fi-plug)  -- (ap);
      \draw [wireless] (camera-1)    -- (ap);
      \draw [wireless] (camera-2)    -- (ap);
      \draw [wireless] (alexa)       -- (ap);

      \draw [wired]    (hue-bridge)  -- (ap);
      \draw [wired]    (st-hub)      -- (ap);
      \draw [wired]    (ap)          -- (router);
      \draw [wired]    (router)      -- (internet);

      \draw [zigbee]   (light-bulb)  -- (hue-bridge);
      \draw [zigbee]   (st-plug)     -- (st-hub);
      \draw [zigbee]   (door-sensor) -- (st-hub);
      \draw [zigbee]   (presence-sensor) -- (st-hub);

      \node (rect-origin) at (2.5,-0.7) {};
      \node (legend)  [below of=rect-origin,xshift=0.6cm,yshift=1.7cm] {\scriptsize Legend};
      \draw [ultra thin] (rect-origin) rectangle (4.3,-2.4);
      \node (wired-a)   [below=0.1cm of legend.south west] {};
      \node (wired-b)   [right=0.6cm of wired-a] {};
      \draw [wired]     (wired-a) -- (wired-b);
      \node (wired-key) [right=0.4cm of wired-a.east] {\tiny Ethernet};
      \node (wireless-a) [below=0.3cm of wired-a] {};
      \node (wireless-b) [right=0.6cm of wireless-a] {};
      \draw [wireless]   (wireless-a) -- (wireless-b);
      \node (wireless-key) [right=0.4cm of wireless-a.east] {\tiny Wi-Fi};
      \node (zigbee-a) [below=0.3cm of wireless-a] {};
      \node (zigbee-b) [right=0.6cm of zigbee-a] {};
      \draw [zigbee]   (zigbee-a) -- (zigbee-b);
      \node (zigbee-key) [right=0.4cm of zigbee-a.east] {\tiny Zigbee};

  \end{tikzpicture}
  \vspace{-2mm}
  \caption{Experimental Smart Home network. The numbers refer to the IDs of the device in Table~\ref{tab:devices}.}
  \label{fig:network}
\end{figure}

\subsection{Traffic capture}
\label{sec:traffic-capture}

We capture the traffic from and to the devices. We use the resulting datasets for the creation of the device profiles as well as for the experiments described in Section~\ref{sec:fuzzing}.

To collect network traffic, we execute \texttt{tcpdump} \cite{tcpdump} on the four network interfaces of the LAN AP,
namely the 2.4 GHz and 5.0 GHz WLAN, the wired LAN, and the WAN interfaces. For the recorded traffic to be as exhaustive as possible with respect to the device's operation, we interact with the devices. In addition to physical interactions, we also communicate with all devices through their respective vendor-provided companion app running on a mobile phone, first connected to the same LAN as the device, then to an external network. Device-specific control commands are issued, e.g., switching the smart plug on and off, or streaming the video camera in real time. For the devices compatible with the Amazon or SmartThings hubs, a similar interaction scheme is applied through the respective hubs' apps and through Amazon Alexa's voice control.
Performing those interactions, we made sure to cover various types of device behavior,
such as derived in \cite{oconnor_homesnitch_2019}.
Finally, we also left the devices idle for some time to capture background traffic and device activities not linked to user activities, such as checks for software updates.

\subsection{Profile creation}
\label{sec:setup-profiles}

We manually build profiles for the devices based on the traffic captures. Since the Philips Hue lamp and the SmartThings end devices (a door sensor, a Smart plug, and a presence sensor) are not IP devices, their behavior is described in the profile of the gateway they are connected to, that is the Philips Hue Bridge and the SmartThings Hub, respectively.
We use the Amazon Echo Dot only to control the other devices and therefore did not create a separate profile for it. However, it \emph{does} appear in the profiles of the other devices as part of the description of their interactions with it.

The profiles are then automatically translated to script and C code files by our translation tool, as described in Section~\ref{sec:system}.
Table~\ref{tab:devices} gives metrics concerning the number of different policies, interactions, and NFQueue queues for the devices which will be instrumented in Section~\ref{sec:evaluation}.

\begin{figure}
  \centering
  \includegraphics[width=\linewidth]{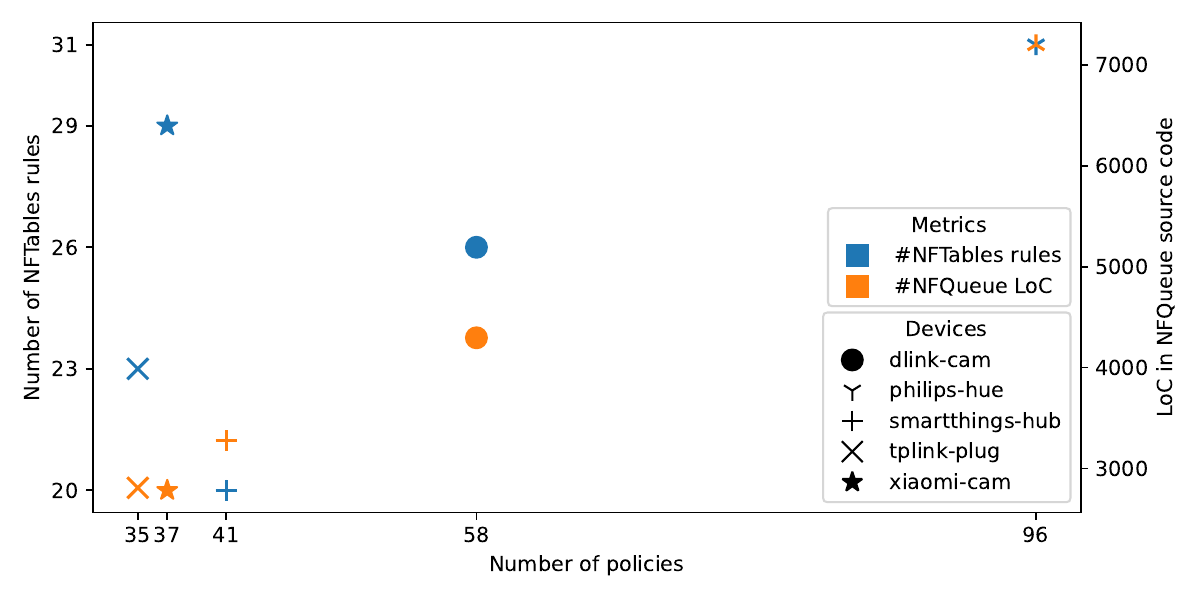}
  \caption{Firewall size metrics per profile size}
  \label{graph:firewall-size}
\end{figure}

Figure~\ref{graph:firewall-size} shows, for each device, the relation between the number of policies in its profile and the number of generated NFTables rules (left y axis) as well as the number of lines of code (LoC) in the corresponding NFQueue source code file (right y axis). The most ``complex'' device among the devices used in the experiments is the Philips Hue bridge and the attached lamp. Unlike a Smart plug, the lamp allows different types of interactions (e.g., adjusting the brightness, the color, etc.). Moreover, the bridge has more connectivity options (mDNS, SSDP) than simpler devices and can directly communicate with devices such as the SmartThings hub without going through a Cloud connection. This results in a greater number of NFTables rules for the different protocols and in a larger NFQueue program to process the interactions. Another notable outlier is the Xiaomi camera: its profile describes interactions with various hosts, which require individual NFTables rules.

\section{Evaluation}
\label{sec:evaluation}

In our evaluation,
we show using packet fuzzing that the firewall indeed blocks interactions that do not match the profile. 
Then, we measure the time taken by our firewall to accept valid packets compared to the baseline without firewall.
Finally, we demonstrate the ability of our approach to protect against attacks.

\subsection{Fuzzing for system validation}
\label{sec:fuzzing}

Our first evaluation step is to validate our system
by verifying that it accurately accepts traffic which conforms to the configured profiles,
whilst dropping any traffic which does not conform.
Non-conforming traffic can take two forms:
on the one hand, packets which have characteristics not complying
with any packet specified in the profile,
either due to their packet signature or their traffic statistics;
on the other hand, packets which comply to the profile,
but are not received in a state where they should be accepted.

To produce such traffic, we developed a fuzzer which takes a packet capture as input,
randomly picks packets, and modifies them to produce a new packet capture.
More precisely, the modification is always applied to the highest supported protocol layer
(i.e., the DNS layer for a DNS packet, the UDP layer for a plain UDP packet, etc.).
Given that the highest layers protocols are sent to the user-space NFQueue program,
and the lowest layers ones (mostly TCP, UDP, and ARP) are directly matched by the NFTables firewall,
this ensures both the packet matching capabilities of NFTables and NFQueue are exerted and tested for correctness.

It must be noted that,
whereas modified packets should be dropped simply because they do not comply to the profile,
unmodified packets might also be dropped
if a previous packet from the same interaction was modified and dropped.
Indeed, this would effectively remove an expected packet from the interaction,
which, depending on the nature of the policy,
might provoke cascading drops for all subsequent packets.

Our testing methodology is the following:

\textbf{Base PCAP creation:} We split the base device traffic described in Section~\ref{sec:traffic-capture} to isolate specific interaction scenarios. This forms our baseline PCAP files. The firewall should not block any of the packets in those files.


\textbf{Fuzzing PCAP files:} We run the aforementioned packet fuzzing program on each PCAP file an arbitrary number of times (here, five) to produce different versions of modified traffic. If a packet part of an interaction has been edited, it could potentially make all the forthcoming interaction packets being dropped, as the interaction pattern will not be respected anymore.

\textbf{Replaying PCAP files:} We replay the edited PCAPs through the firewall, and log information for each packet, including its verdict, the policy, and interaction it belongs to.

\textbf{Labeling packets:} We label packets with their expected verdict:
If the packet was edited and does not conform to the device profile anymore, its expected verdict is DROP.
If the packet was \emph{not} edited, we must check if a previous packet in the same interaction was edited. If yes, and if this should prevent the current packet from being accepted, the expected verdict is DROP. Otherwise, the expected verdict is ACCEPT.

We replayed a total of 10894 packets.
Among them, 4348 were supposed to be accepted,
and 6546 to be dropped.
Our firewall successfully issued the correct verdict for all packets,
and therefore showcases perfect precision and accuracy scores.

\subsection{Latency induced by the firewall}
\label{sec:latency}

As our system is intended to inspect live traffic,
it must be transparent to the user of the Smart Home network,
and not hinder the correct operation of any device in the network.
As such, the latency induced by the firewall in the network must stay negligible.
With this intent, we now evaluate our system's added latency on the network's AP packet forwarding.
To this end, we deploy the testbed network presented in Section~\ref{sec:testbed-devices}
and shown in Figure~\ref{fig:network}.
We evaluate the latency induced on the network traffic related to five of our devices, namely:
\begin{itemize}
    \item the D-Link camera \cite{dlink-cam}
    \item the Philips Hue bridge \cite{hue-bridge}, connected to one color lamp \cite{color-lamp}
    \item the SmartThings hub \cite{smartthings-hub}, connected to three Zigbee devices: a smart plug \cite{st-plug}, a door sensor \cite{multi-sensor}, and a presence sensor \cite{presence-sensor}
    \item the TP-Link smart plug \cite{hs110}
    \item the Xiaomi camera \cite{xiaomi-cam}
\end{itemize}

As explained in section \ref{sec:setup-profiles}, the behavior of the Zigbee devices (namely the Philips Hue lamp, and the SmartThings plug, door sensor, and presence sensor) is embedded in the profile of their related gateway (Philips Hue Bridge or SmartThings Hub),
whereas the Amazon Echo Dot is only used to control the devices,
and does not have a specific profile.

First, we activate one device at a time and interact with it.
For each device, we record traffic on the four AP network interfaces.
We compute the latency induced by processing on the AP,
by taking the difference between the packet timestamp on the egress and ingress interfaces.
We repeat this measurement in three different scenarios:
\begin{itemize}
    \item \textbf{no-firewall}: without any firewall rule active.
    \item \textbf{base-firewall}: with the default built-in OpenWrt firewall rules active on the AP. This configuration is not intended to block any attack. 
    \item \textbf{my-firewall}: with our firewall active for the device under test.
\end{itemize}
In the \textbf{my-firewall} test, our firewall processed 2105 packets (1.7 MBytes) for the D-Link camera (during 10 minutes, of which the camera was streaming during 20 seconds), 3862 packets (0.8 MBytes) for the Philips Hue,
1071 packets (0.2 MBytes) for the SmartThings hub,
682 packets (0.2 MBytes) for the TP-Link plug,
and 31816 packets (17.2 MBytes) for the Xiaomi camera (during 30 minutes, of which the camera was streaming during 120 seconds).

Finally, in the last scenario (called \textbf{all-devices}), we activate and interact with the five devices at the same time to mimic a more realistic Smart Home network. In this test, 32k packets (17.1 MBytes) were exchanged during 53 minutes.

\begin{figure}
    \centering
    \includegraphics[width=0.9\linewidth]{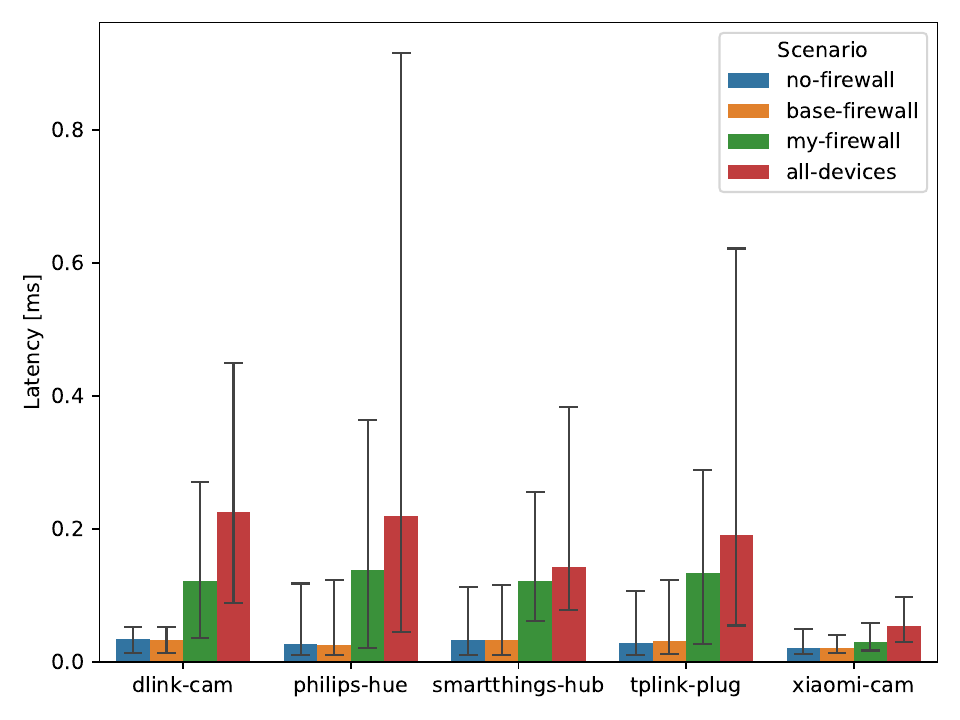}
    \caption{Mean latency induced by the AP hosting the firewall, per experimental scenario}
    \label{graph:latency}
\end{figure}

Figure~\ref{graph:latency} shows the mean and 95-percentile range (the interval from the 2.5 to the 97.5-percentile) of the latency in the four scenarios for the five exerted devices.
We observe that, for all devices, the 95-percentile range of the processing latency is under one millisecond,
even when our firewall is active.
This latency is negligible compared to the network latency,
which is typically of the order of 2 to 5 ms between devices in the same LAN,
i.e., the fastest type of communication we cover.
Moreover, typical human reaction times are always over the millisecond \cite{attig_system_2017}.
We can therefore conclude that the latency induced by our firewall
is not an issue and will not be noticed by the user when interacting with the devices. 

It must also be noted that
in the scenario where all devices are connected to the network
and their corresponding firewall rules are all active at the same time,
the firewall-induced latency increases.
This effect is expected, as the system must process more packets and rules.
However, the increase stays at an acceptable, user-imperceivable level.
This indicates that our system is able to accommodate multiple devices
in the context of a Smart Home network,
where the number of devices hardly exceeds ten \cite{huang_iot_2020}.


We further investigate the origin of the latency induced by the firewall
by splitting the processed packets in four categories,
ranked in increasing firewall processing time:

\begin{itemize}
    \item \textbf{A}: packets which were matched by plain NFTables, without queueing to NFQueue;
    \item \textbf{B}: packets which were queued to NFQueue and required string comparison;
    \item \textbf{C}: packets which were queued to NFQueue and required domain name lookup in a DNS map;
    \item \textbf{D}: packets which were queued to NFQueue and underwent string comparison \emph{and} domain name lookup in a DNS map;
\end{itemize}

\begin{figure}
    \centering
    \includegraphics[width=0.9\linewidth]{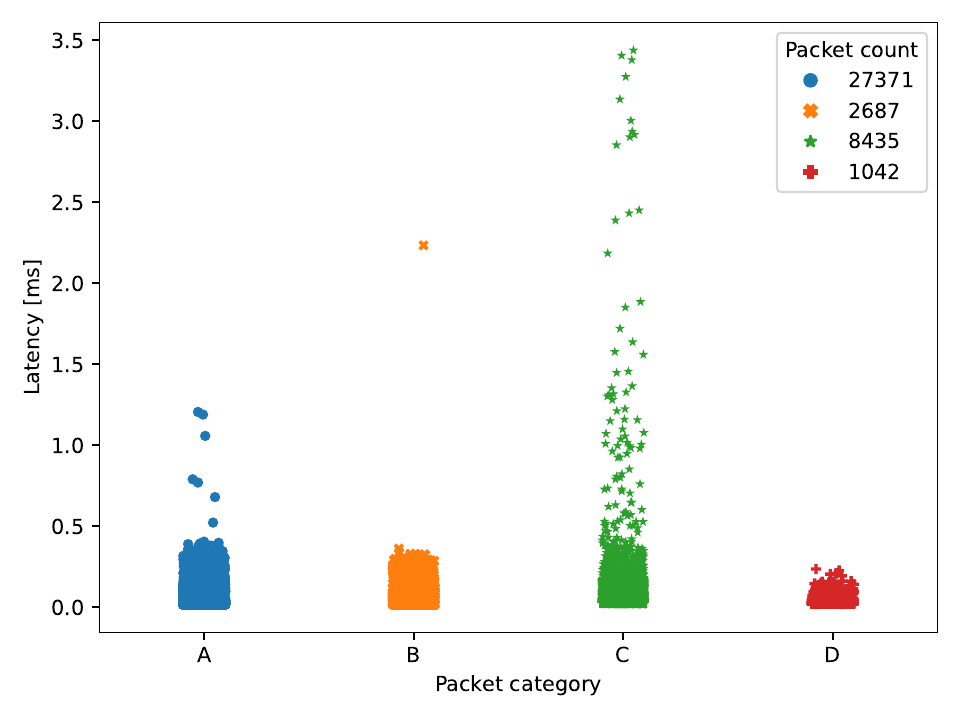}
    \caption{Firewall latency per processing effort category}
    \label{graph:packet-count}
\end{figure}

We then count the number of packets corresponding to each category,
and plot the latency for each category in Figure~\ref{graph:packet-count}.
We observe the joint effect of the sheer packet number and the category on the latency.
Indeed, category \textbf{A} is the less time-consuming one,
but the fact that it has a lot of corresponding packets
increases its total latency above categories \textbf{B} and \textbf{D}.
Furthermore, the most time-consuming category is \textbf{C},
due to its large number of packets \emph{and}
its inherent processing complexity,
triggering the user-space NFQueue program
instead of only matching packets with NFTables. \textbf{D}-typed packets 
require more processing than \textbf{C}-packets but are less numerous, and thus 
less delayed in average.

\subsection{Capability to block attacks}


As argued in Section~\ref{sec:threat-model}, our approach is, by design, limited to attacks that have, directly or indirectly, an impact on the activities in the network. Such attacks include reflection attacks, e.g., by DNS \cite{anagnostopoulos2013dns} or NTP \cite{czyz2014ntp}, and other types of brute-force DoS attacks, network scans, data exfiltration, attempts to remotely control a device from an unknown host, send unexpected messages, etc. 
We do not show evaluation results for all of these attacks since many of them are trivially blocked because they use addresses, ports, or protocols that are not defined in the profiles.

In this section, we take back and implement the three attacks
described in Section \ref{sec:attack-examples}.
As a reminder, each attack stresses a different capability of our system,
respectively the principle of device interactions, the traffic statistics,
and the matching for new, higher-layer protocols.

We then ran the attacks while our firewall was active,
and recorded the accepted and dropped packets,
to determine if the system was able to block the attacks.
Besides, similarly to the firewall's latency evaluation presented in Section~\ref{sec:latency},
we computed the latency induced by our system when the network undergoes the attack,
to compare it to the case without attacks.

In the remaining of this section,
we will, for each attack, detail its implementation
and the outcome of the firewall,
i.e. which packets were accepted or dropped.
Afterwards, we will compare the latency induced by the firewall
in the cases without and with attacks.

\subsubsection{Attack A - Device Interaction}
\label{sec:attack_interaction}

The first attack was designed to illustrate the concept of device interaction.
The test interaction was implemented using the SmartThings hub and its attached door sensor,
as well as the TP-Link smart plug.
We assume the intended behavior of the network is the following:
\begin{enumerate}
    \item The hub, upon detection of change in the door sensor's state, transmits a message to the vendor's cloud, indicating this change.
    \item The cloud processes the message, and instructs the smart plug to toggle, by sending it the required message.
\end{enumerate}

The attack script purposefully transmits network patterns corresponding to the second step of the interaction,
without the first step having been seen.
This simulates an attacker trying to issue an unwanted command to the plug,
whereas its trigger, i.e. the door opening or closing,
has not happened.
This attack was actually split in two versions,
differing in the IP address of the attacker,
mimicking an attacker located either in the same local network,
or remotely.

As expected, the firewall blocks the attack packets,
as the preliminary pattern had not been seen beforehand.

\subsubsection{Attack B - Traffic Statistics}
\label{sec:attack_stats}

The second attack exerts the added matching on traffic statistics.
We added a simple rule, limiting the rate of ARP requests from
a smartphone present in the local network towards the plug to one packet per second.
The attack script then simply issues such requests with a higher rate,
namely 2 packets per second.

\subsubsection{Attack C - new protocols matching}
\label{sec:attack_protocols}

The third and last attack exhibits newly added matching on
previously unsupported protocols,
namely the DNS protocol as example.
Whereas DNS queries from the plug are allowed for given domain names
(i.e. NTP servers, manufacturer's cloud, etc.),
the attack's implementation consists in issuing DNS queries for
unintended domain names.
This mimics a scenario where an attacker would have compromised the device,
and instructs it to communicate with a malicious server.

Contrarily to the base version of MUD,
our system is able to detect this unintended domain name,
and to block the corresponding packets,
effectively preventing an attacker to contact a malicious server in this way.

\subsubsection{Firewall latency under attack}

\begin{figure}
    \centering
    \includegraphics[width=0.9\linewidth]{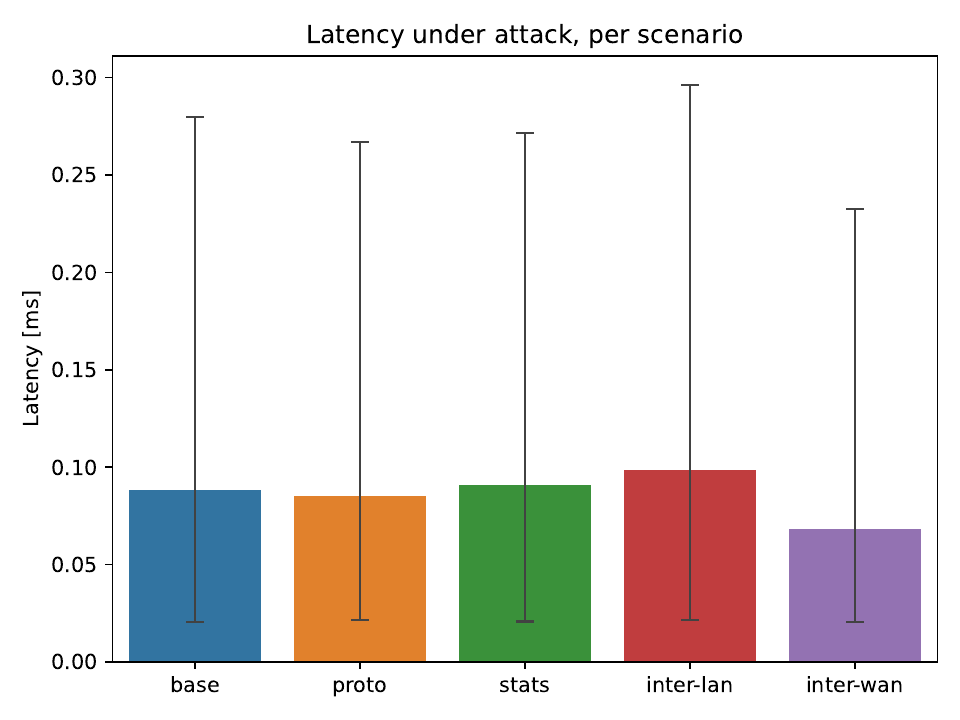}
    \caption{Comparison of the firewall's latency, without and with attacks}
    \label{graph:latency-attacks}
\end{figure}

Finally, we measure the latency induced by our protection system,
when each of our attack scripts is active,
and compare the results to the base case
when there is only benign traffic.
Figure~\ref{graph:latency-attacks} shows the mean and 95-percentile range of the firewall latency,
in the cases without and with attacks.

In all cases,
the latency without and with attack is nearly identical.
The effect of the attack on the latency can be considered as insignificant.
We can thus conclude that our system is resilient to attack traffic.

\subsection{CPU and Memory usage}

For this last evaluation,
we benchmarked the CPU and memory usage of our system
under normal and extreme conditions,
with the incentive of showcasing its robustness
and processing capabilities in case of higher network load.
As such, we deployed the TP-Link smart plug,
along with our system configured for this device.
We then evaluated the aforementioned metrics in four scenarios,
with increasing computational resources requirements:
\begin{itemize}
    \item \textbf{Normal}: Normal operation, i.e., identical to the latency measurements presented in Section~\ref{sec:latency}
    \item \textbf{State}: Under a steady flow of packets rejected by NFQueue because of the network FSM \emph{state}. The firewall must only check the FSM state to issue the reject verdict.
    \item \textbf{String}: Under a steady flow of DNS requests with an unexpected domain name. The firewall must perform string comparison to issue the reject verdict.
    \item \textbf{Lookup}: Under a steady flow of packets directed towards an IP address not present in the DNS table. The firewall must perform a lookup in the DNS table to issue the reject verdict.
\end{itemize}
For the last three scenarios, the stress packet rate was fixed at one per millisecond.

\begin{figure}
    \centering
    \includegraphics[width=0.9\linewidth]{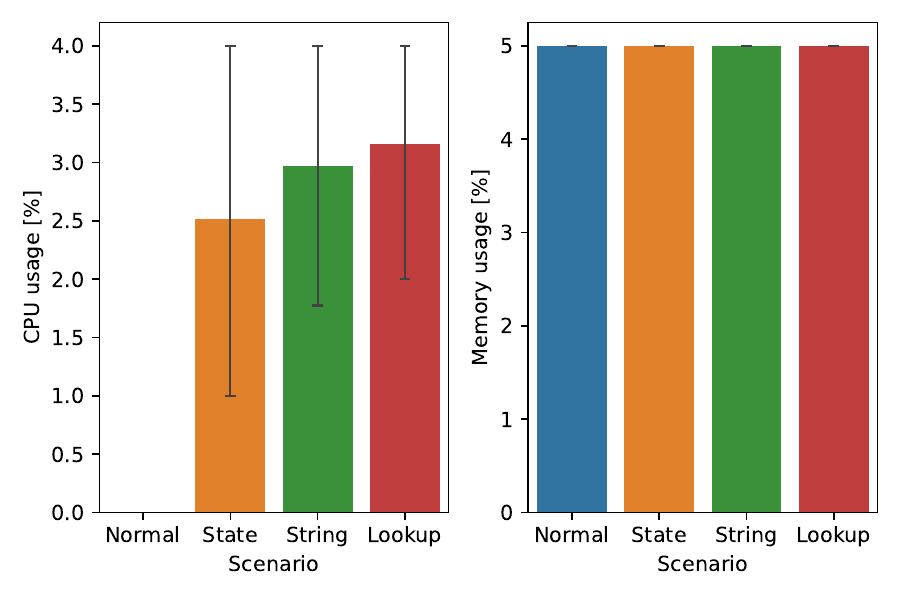}
    \caption{CPU and Memory usage percentage per scenario}
    \label{graph:cpu_metrics}
\end{figure}

Figure~\ref{graph:cpu_metrics} shows the mean and the 95-percentile range for the CPU and memory usage percentage,
for each of the four scenarios.
We observe that both stay very low,
i.e., a maximum of 4\% for the CPU usage,
and 5\% for the memory usage,
even on modest hardware.
This shows our tool can handle traffic loads which are considered high for Smart Home IoT networks.
Additionally, it is very lightweight and can be deployed on virtually any home router supporting OpenWrt.

\section{Discussion}
\label{sec:discussion}

\subsubsection*{Types of attacks that can be blocked or mitigated}

As explained, our approach focuses on attacks with an observable impact on the network behavior of the affected devices.
Since the firewall monitors all its interfaces, it can also block attacks \emph{from} Smart Home devices, such as outgoing DDoS attacks or scans. 
Our firewall performs a limited form of deep packet inspection, for example to match the protocol fields of DNS and HTTP traffic. Apart from that, payload is not further analyzed, which means that traffic that does not look suspicious on metadata level or statistics level (often called \emph{semantic} attacks) cannot be stopped. An example for that would be a compromised firmware downloaded from the manufacturer's server.

Data exfiltration attacks, which break the privacy of a Smart Home, can be blocked depending on their nature. Apart from blocking data transfers to unknown destinations, the firewall can be used to limit the rate, duration, and size of network activities. Therefore, even if the destination is generally trusted (e.g., the server of a manufacturer), the firewall can block data transfers that deviate from the expected behavior, such as a Smart Speaker with a microphone sending more data than expected.

\subsubsection*{Visibility of network traffic}

Our prototype implementation of the firewall is currently limited to WiFi traffic. Some Smart Home devices, like the tested Philips Hue lamp, use Personal Area Network (PAN) technologies, such as Zigbee, 6LoWPAN, or Bluetooth, and connect to the rest of the world through a gateway. If such a device communicates with the Internet or with another device connected directly or indirectly to the network monitored by the firewall, our firewall will see the communication and is able to block it.
By its nature, the firewall cannot act on ad-hoc communication between devices or on communication using long-range networks, such as 4G or LoRa. To monitor and block illegitimate interactions over a mix of access technologies without a central point requires a collaborative firewalling solution implementing our profiles and sharing state of past communications. 

\subsubsection*{Profile creation}

As the starting point of our system are the device profiles, it is important that those are retrieved from a trusted source, and accurately and exhaustively describe the devices' communication patterns.
In the context of this research, we supposed profiles would be created and distributed by the devices' manufacturers (an assumption identical to MUD's) or a group of experienced users. The secure distribution of trusted profiles entails some challenges which we have already discussed in Section~\ref{sec:threat-model} and which we consider to be outside the scope of this paper.

Other strategies to create device profiles can be imagined. Inspired by the work of Hamza \textit{et al.} \cite{hamza_clear_2018} for MUD profiles, one could design a tool that automatically creates our profiles from traffic captures. Since our profile language is much more complex than MUD, we believe that this deserves more research, which we deem as future work. Efforts in this direction, although not specific to our profile language, have been described in the existing literature (see Section~\ref{sec:related-work}). However, it is worth considering whether a \emph{completely automated} creation of profiles based on traffic captures is desirable at all since there is a risk that unwanted or harmful behavior could be inadvertently included in a profile. 

\subsubsection*{Device and profile management}


Another question is how the firewall is provided with the profiles needed for a concrete Smart Home installation.
A simple solution would be to extend our system by a GUI where the user can select and identify the devices they own.
For a gateway, its profile is obtained by merging the profiles of the individual devices attached to it.
The firewall could then download the necessary profiles from a trusted repository. Alternatively, the firewall could try to automatically identify the devices present in the network. Again, we consider this to be outside the scope of this paper, as numerous research works have been already conducted on device identification \cite{sivanathan_classifying_2019, oconnor_homesnitch_2019, saidi_haystack_2020}.

Finally, it should be noted that profiles might need to be updated during the lifetime of the firewall. During our experiments we had to do this for one of the devices after it received a firmware update that changed its behavior.

\section{Related work}
\label{sec:related-work}

Existing related work can be divided into three areas: The usage of device profiles to enforce security policies, protection systems for IoT networks in the context of Smart Homes, and the extension of MUD.

\subsection{IoT device profiling}

In an effort comparable to the MUD standard and our work, Barrera \textit{et al.} \cite{barrera2018standardizing} present a methodology to derive network behavior profiles for Smart Home devices from traffic captures. The profile of a device contains information about the network communication it is allowed to do. A fundamental limitation of their work is the fact that their profiles only describe basic characteristics of the expected network communication, e.g., destination IP addresses, protocol types, and rate limits. This limitation allows a straight-forward translation of the profiles to IPTables rules in order to block unwanted traffic.
In our work, profiles are much more expressive and can be used to describe complex interactions between devices. Since this is beyond the capabilities of simple IPTables rules, we follow a different approach to traffic filtering, which we have described in detail in the previous sections.

Hamza \textit{et al.} \cite{hamza_clear_2018} developed the \textit{MUDgee} tool,
which can produce the MUD profile corresponding to a device
based on traffic captures.
As the created profiles comply with the base MUD specification,
they suffer from MUD's shortcomings,
which we have alleviated with our new profile specification.

Hu \emph{et al.} recently proposed BehavIoT \cite{hu_behaviot_2023},
a tool which can extract models of the behavior of a complete Smart Home network.
More precisely, it inspects the network traffic to fingerprint and classify traffic in three categories:
periodic events, which represent necessary background traffic to other hosts, mainly cloud servers or the local gateway;
user events, triggered when a user issues an action towards a device in the network (e.g., turning on a light bulb);
aperiodic events, the remaining ones, which have been analyzed to usually not be necessary for the network's correct operation.
Their work is flexible enough to be able to identify network patterns resulting from device interactions,
as described in section \ref{sec:device-interactions},
additionally to single device patterns.

It is conceivable to use the different tools proposed by the above authors to automatically create the device profiles described in our paper, but, as we already mentioned in Section~\ref{sec:discussion}, the question of how to ensure that undesirable behavior is not learned by such tools, must then be addressed.

\subsection{IoT and Smart Home protection systems}

Numerous authors have developed security systems for IoT or Smart Home networks.
The vast majority leverages Machine Learning (ML) models.

A typical example of such ML-based systems is \textit{Passban IDS}, developed by Eskandari \textit{et al.} \cite{eskandari_passban_2020}. It is an anomaly-based IDS which can be deployed on gateways with commodity hardware. Another example following the same general approach, namely network-based and anomaly-based detection, is proposed by Pashamokhtari \textit{et al.} \cite{pashamokhtari_progressive_2020}. Their system is intended for use in a Smart Home network and employs sequential ML models of increasing complexity that are specialized in a subset of traffic features to describe the expected traffic of IoT devices. In contrast to our and Eskandari's work, they employ Software Define Networking (SDN) to dynamically adapt the traffic forwarding rules.
 
Oliva and Mohandes \cite{oliva2022smart} did not use ML-techniques but implemented a stateful Smart Home firewall leveraging, like us, NFTables and NFQueue. However, their work focuses on two widespread IoT application-layer protocols, namely CoAP and MQTT, whilst we support various types of protocols on multiple layers.

Hamza \textit{et al.} \cite{hamza_combining_2018} were the first to leverage MUD profiles
to provide network security to live traffic. They did so by translating the MUD profiles' ACLs into SDN rules.
Traffic complying with those rules is accepted, whilst other traffic is forwarded to an instance of the Snort IDS \cite{roesch1999snort}. Our firewall is designed to run on commodity hardware commonly found in Smart Homes and does not target SDN-capable equipment or hardware able to run the Snort IDS.

\subsection{Extensions to the MUD standard}

As discussed in Section~\ref{sec:attack-examples}, the MUD standard suffers from multiple shortcomings.
Existing works have tried to overcome those shortcomings, or to add new capabilities to MUD,
by developing extensions to the standard.

Lear and Friel \cite{lear-bandwidth-2019} proposed ``bandwidth extensions'', which allow matching on the traffic rate.
Our profile language supports rate limitation as well as new matchable traffic statistics such as size and duration. However, in our language, they are integrated into the interaction mechanism described in Section~\ref{sec:interactions} and have therefore an extended meaning beyond simple packet matching.

Reddy \textit{et al.} \cite{reddy_mud_2023} developed an extension providing parameters to match on TLS features, e.g., encryption, signature and key exchange algorithms. Matheu \textit{et al.} \cite{matheu_mud} extend MUD with the capability of matching application layer protocols by their name, as well as the number of concurrent connections using a specific protocol and the accessed resources.
We have left out support for TLS feature matching in our prototype and focus on application layer protocols commonly found in Smart Home device communication. Whilst contrary to Matheu \textit{et al.} \cite{matheu_mud}, we provide protocol-specific matching capabilities and not just protocol name matching.

Morais and Farias \cite{morais_inxu_2020} added \emph{Malicious Traffic Description} fields (MTD) to MUD profiles in an effort to include signature-based protection by matching and blocking known attacks.
Finally, Matheu \textit{et al.} \cite{matheu_security_2020} leverage the Medium-level Security Policy Language (MSPL) \cite{zarca2019security} to provide data privacy, resource authorization, and channel protection capabilities to MUD.

\section{Conclusion}
\label{sec:conclusion}

Modern Smart Home installations exhibit complex communication patterns caused by the direct and indirect interactions of the deployed devices. For ordinary users, it becomes difficult to configure security components such as firewalls so that they allow legitimate network activities and block unwanted ones.

Taking inspiration from the IETF's MUD standard, we defined a new language for communication profiles of IoT devices. Contrary to MUD, the language is expressive enough to describe communication patterns involving multiple devices and their interactions. 
We designed a lightweight firewall that uses the profiles to enforce communication policies in Smart Home networks. We performed experiments with various types of IoT devices and showed that the firewall is effective in blocking unwanted network activities, while remaining transparent to the devices and the users.

As future work, we envision the secure automatic creation of interaction profiles from network captures and extending the approach to other types of networks.

\section*{Acknowledgements}
F. De Keersmaeker is funded by the F.R.S.-FNRS Research Fellow [ASP] grant with number 40006064.
F. De Keersmaeker, R. Sadre and C. Pelsser are affiliated with UCLouvain, Belgium.
All icons are from \rurl{flaticon.com}.

\bibliographystyle{plain}
\bibliography{biblio}

\appendices
\newpage

\section*{Appendix}
\label{sec:appendix}

\subsection*{Larger device profile example}
\label{sec:larger-profile}

\begin{figure}[H]
    \centering
    \includegraphics[width=0.9\linewidth]{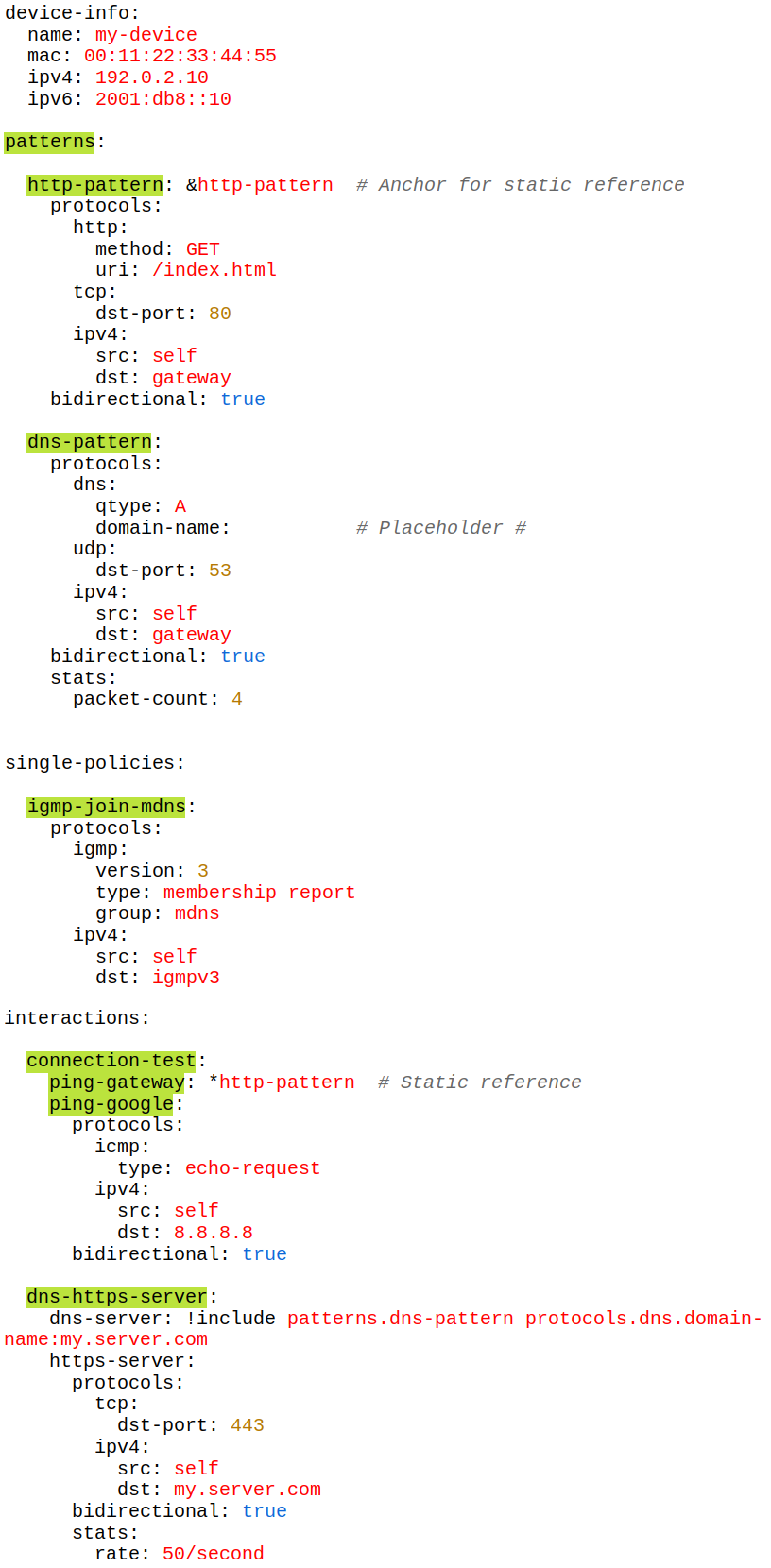}
    \caption{
        Larger device profile example.
        User defined labels are highlighted in green. Builtin keywords are shown in black.
    }
    \label{fig:profile-full}
\end{figure}

Figure~\ref{fig:profile-full} shows a more complete device profile example.
It illustrates the following components:
\begin{itemize}
    \item Static references to fields in the \textbf{same profile}. The original field is tagged with the \texttt{\&} symbol and a name, then the referencing field uses the \texttt{*} symbol and the same name.
    \item References to fields in any profile, with the \texttt{!include} directive. This construct allows to set a value for a given field in the referenced pattern. For instance, in this example, the field \texttt{domain-name} in the \texttt{dns-pattern} object will be filled with the value \texttt{my.server.com}.
    \item Easy matching of bidirectional traffic with the \texttt{bidirectional} keyword.
    \item Multiple matching capabilities, including various protocols and traffic statistics, namely packet count and rate.
\end{itemize}

\newpage

\end{document}